\documentclass[english]{emulateapj}
\usepackage[T1]{fontenc}
\usepackage{amstext}
\usepackage{graphicx}

\makeatletter
\slugcomment{}
\shorttitle{}
\shortauthors{}

\makeatother

\usepackage{babel}
\begin{document}

\title{An adjoint based method for the inversion of the Juno and Cassini
gravity measurements into wind fields }

\author{Eli Galanti}

\affil{Weizmann Institute of Science, Rehovot, Israel}

\email{eli.galanti@weizmann.ac.il}

\and{}

\author{Yohai Kaspi}

\affil{Weizmann Institute of Science, Rehovot, Israel}

\begin{abstract}
During 2016-17 the Juno and Cassini spacecraft will both perform close
eccentric orbits of Jupiter and Saturn, respectively, obtaining high-precision
gravity measurements for these planets. This data will be used to
estimate the depth of the observed surface flows on these planets.
All models to date, relating the winds to the gravity field, have
been in the forward direction, thus allowing only calculation of the
gravity field from given wind models. However, there is a need to
do the inverse problem since the new observations will be of the gravity
field. Here, an inverse dynamical model is developed to relate the
expected measurable gravity field, to perturbations of the density
and wind fields, and therefore to the observed cloud-level winds .
In order to invert the gravity field into the 3D circulation, an adjoint
model is constructed for the dynamical model, thus allowing backward
integration. This tool is used for examination of various scenarios,
simulating cases in which the depth of the wind depends on latitude.
We show that it is possible to use the gravity measurements to derive
the depth of the winds, both on Jupiter and Saturn, taking into account
also measurement errors. Calculating the solution uncertainties, we
show that the wind depth can be determined more precisely in the low-to-midlatitudes.
In addition, the gravitational moments are found to be particularly
sensitive to flows at the equatorial intermediate depths. Therefore
we expect that if deep winds exist on these planets they will have
a measurable signature by Juno and Cassini.
\end{abstract}

\section{Introduction}

\begin{figure}
\begin{centering}
\includegraphics[width=1\columnwidth]{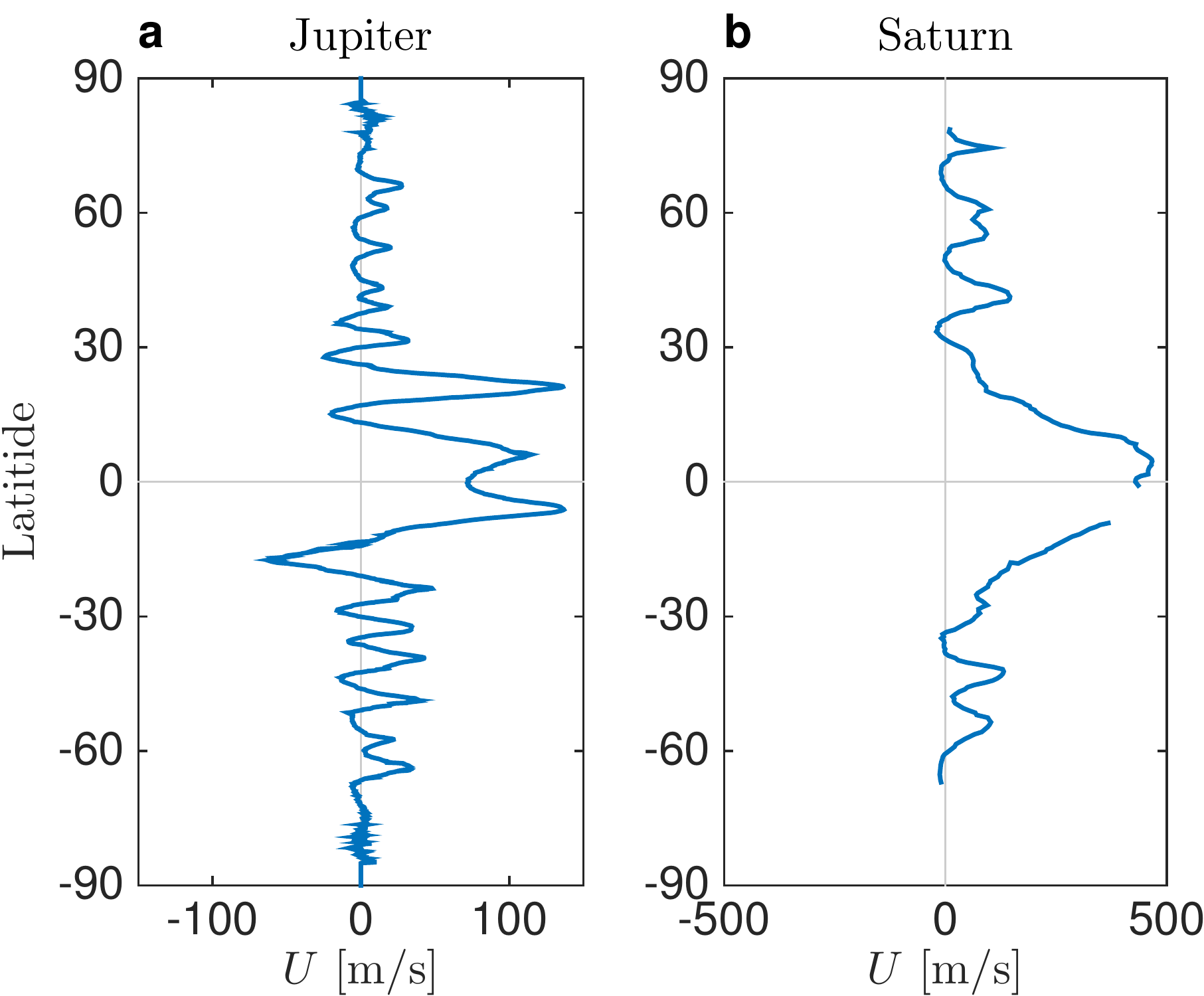}
\par\end{centering}

\caption{(a) Surface winds on Jupiter \citep{Porco2003}, and (b) surface winds
on Saturn. \citet{Sanchez-Lavega2000}\label{fig:winds}}
\end{figure}

At the observed cloud-level of both Jupiter and Saturn their atmospheric
dynamics are dominated by strong east-west (zonal) jet streams (Fig.~\ref{fig:winds}),
reaching velocities of $140\,\textrm{ms}^{-1}$ on Jupiter and over
$400\,\textrm{\text{ms}}^{-1}$ on Saturn \citep{Vasavada2005}. It
is currently unknown how deep these jets extend \citep[e.g.,][]{DelGenio2012,Li2006},
and the only available direct measurements below the cloud-level are
from the 1995 Galileo probe to Jupiter that found $160\,\text{ms}^{-1}$
winds extending down at least to $22\,\text{bars}$ at the entry point
of the probe ($6^{\circ}$N) \citep{Atkinson1996}. Addressing this
question is one of the main goals of the Juno mission to Jupiter and
the Cassini proximal orbits at Saturn, aiming to determine the depth
extent of atmospheric circulation on these planets through precise
measurements of their gravity field \citep{Hubbard1999,Kaspi2010a}.
This will possibly allow answering the long lasting debate regarding
the depth of the dynamics on the giant planets, and thus shed light
on the mechanisms the could be driving the jets \citep[e.g.,][]{Busse1976,Williams1978,Cho1996,Showman2006,Scott2007,Kaspi2007,Lian2010,Liu2010,Liu2013}.

The Juno mission was launched in 2011 and will arrive at Jupiter in
2016 equipped to perform high precision measurements of the gravity
field with expected accuracy that will allow meaningful measurements
up to at least $J_{12}$ \citep{Bolton2005}. In 2017, NASA's Cassini
mission will conclude its 13 year tour of the Saturnian system, with
planned proximal orbits of Saturn obtaining the same type of data
for Saturn, just before the spacecraft terminates its operation by
descending into Saturn's interior. For both spacecraft the detection
of the gravity signal will be done by Doppler tracking of the spacecraft
trajectory.

In recent years, in anticipation of the arrival of Juno at Jupiter,
several studies have looked at the effect of interior flow on the
gravitational signature of the planet. To leading order, the gravity
spectrum is affected by the planet's oblate shape and radial density
distribution. However, on giant gas planets, since the planet is composed
mainly of light elements, and have no solid surface, the relative
effect of density perturbations due to their internal and atmospheric
dynamics can be significant and affect the measured gravity field.
Particularly, if the strong winds extend deep enough into the planets'
interior, their relative effect on gravity becomes larger. This was
first noted by \citet{Hubbard1982} and later developed further by
\citet{Hubbard1999}. In these studies, potential theory (the adjustment
of potential surfaces under rotational and internal structure constraints)
was used to show that if differential rotation on Jupiter penetrates
the depth of the planet, then the resulting high-order gravity moments
will be stronger than the corresponding solid-body moments. This approacvh
was recently further developed using a more accurate concentric MacLaurin-based
interior models \citep{Hubbard2012,Kong2012,Hubbard2013}.

These studies have allowed accurate estimation of the gravity field,
but have been limited to flows following full cylindrical symmetry
as these potential theory models are limited to fully barotropic systems
in which the flow is constant along lines parallel to the axis of
rotation. A second approach proposed was using thermal wind balance
models \citep{Kaspi2010a,Kaspi2013a,Kaspi2013c,Liu2013}, where the
gravity field resulting from any given wind field could be calculated,
but these models are limited to spherical symmetry, resulting in inability
to calculate the static (solid-body) gravity spectrum and neglecting
the effect of the planet oblateness on the wind contribution to the
gravity moments. \citet{Kong2012} calculated this effect for the
case of the full barotropic flow, and found it to be small. Similarly,
using a thermal-wind based model with an oblate mean state density
structure we find that the effect of oblateness on the dynamical contribution to the gravity
moments is small \citep{Kaspi2013d,Kaspi2016}.

Because Jupiter and Saturn are gaseous, aside from the cloud-level
winds there is no apparent asymmetry between the northern and southern
hemisphere. Therefore, the gravitational moments resulting from the
shape and vertical structure of the planets have identically zero
odd moments \citep{Kaspi2013a}. However, the observed cloud-level
wind structure does have hemispherical differences, and it was shown
that even if these asymmetries extend only $O(100)$~km below the
surface their contribution to the odd gravity moments is measurable
\citep{Kaspi2013a}. Unlike the even moments that have a contribution
both from the static density distribution and the dynamics, the odd
moments are caused purely due to dynamics. Thus, any odd signal detected
($J_{3},J_{5},J_{7},...$) will be a sign of a dynamical contribution
to the gravity signal, and this might be one of the first signals
of deep dynamics that might be measured by Juno and Cassini. Using
also the thermal wind approach, \citet{Liu2013} calculated the penetration
depth of the winds on Jupiter with the additional assumption that
the entropy gradient in the direction of the spin axis must be zero.
This requirement sets the penetration depth of the winds, and they
have also found that such a wind structure should be detectable by
Juno and Cassini \citep{Liu2014}. 

All studies to date have been only in the direction of forward modeling;
thus, given a hypothetical wind structure (based on the observed surface
winds and some assumption regarding the penetration depth) the gravity
moments are calculated via the effect of the winds on the density
structure \citep[e.g.,][]{Hubbard1999,Kaspi2010a,Liu2013,Kong2012}.
However, in order to analyze the gravity field that will be detected
by Juno and Cassini we need to solve the inverse problem, and calculate
the zonal wind profile \textit{given} the gravity field. This causes
a difficulty since a gravity field is not necessarily invertible,
and a given gravity field might not have a unique corresponding wind
structure. Here, we propose to address this issue using an adjoint
based inverse method that will allow the investigation of the giant
planet dynamics using the observed measured gravity field. This method
has been used extensively in the study of oceanic and atmospheric
fluid dynamics \citep[e.g.,][]{Tziperman1989,Mazloff2010,Moore2011,Kalmikov2014}.
In this study we present results based on the zonal winds only, but
this method will enable to relate to the full gravity maps, not only
for the zonal moments, but for the full 3D gravity fields including
contributions from longitudinal variations in the wind structure and
meridional winds. These variations might be detectable if the depth
of these longitudinal features have a depth of at least few thousands
kilometers \citep{Parisi2016}. Moreover, the adjoint model we present
in this study is derived using the forward model based on the thermal
wind method. However, the inversion method is more general and can
be applied to more complex models (e.g., \citealt{Zhang2015}) or
general circulation models as is often done in ocean science \citep[e.g.,][]{Galanti2003,Mazloff2010}.

In section \ref{sec:Methods} we describe the thermal wind forward
method for calculating the wind induced contribution to the gravity moments, its adjoint
counterpart, and the optimization procedure used to find the depth
of the winds. In section \ref{sec:Results} we discuss the results
for several cases, a case with wind depth that is not varying with
latitude, a case where the winds depth on Jupiter is allowed to vary
with latitude, and the same analysis applied for Saturn. We also discuss
sensitivities to flow perturbations in the planet deep interior. We
discuss the results and their implications for the Juno and Cassini
missions in section \ref{sec:Discussion-and-Conclusion}.

\section{Methods}
 \label{sec:Methods}

The relation between the density structure of the planet and the resulting
gravity signature can be interpreted using the zonal gravity moments,
which are defined as 
\begin{equation}
J_{n}=-\frac{1}{Ma^{n}}\int P_{n}\rho r^{n}d^{3}\mathbf{r},\label{eq: zonal harmonics-1}
\end{equation}
where $M$ is the planetary mass, $a$ is the mean planetary radius,
$P_{n}$ is the $n^{\text{th}}$ Legendre polynomial and $\rho$ is
the local density \citep{Hubbard1984}. The density can be divided
into the solid-body component $\tilde{\rho}\left(r,\theta\right)$,
and a dynamical component $\rho'\left(r,\theta\right)$ arising from
the fluid motion ($\theta$ is latitude), so that $\rho=\tilde{\rho}+\rho'$
(see \citet{Kaspi2013a} for more details). Similarly, in our analysis
we separate the gravity moments to the static gravity signal, which
is due to the static density mass distribution of the planet and is
calculated using an internal structure model, and the contribution
from the dynamical density perturbations due to the zonal flows. As
our main goal is to determine the penetration depth of the observed
zonal flows, we take these as a given and allow a wide range of penetration
depths, which are the parameters we are trying to optimize.

\subsection{The thermal wind forward model}

Starting from the observed cloud-level winds, we first need to establish
the nature of the subsurface flow. Since the planet is rapidly rotating,
and Coriolis accelerations are dominant over the inertial accelerations
(small Rossby number), surfaces of constant angular momentum will
be nearly parallel to the axis of rotation \citep{Kaspi2009,Schneider2009}.
Conservation of angular momentum then implies, to leading order, that the flow is mainly zonal, i.e., that any meridional circulation will be much weaker than the zonal flow. These zonal flows have been shown in numerical models to have a structure which is aligned with the axis of rotation, yet with wind speeds that decay with depth \citep{Kaspi2009}. Therefore, similar to \citet{Kaspi2010a} we assume that the zonal wind field
has the form
\begin{equation}
u\left(r,\theta\right)=u_{0}\exp\left(\frac{r-a}{H}\right),\label{eq:zonal wind field}
\end{equation}
where $u_{0}\left(r,\theta\right)$ are the observed cloud-level zonal
winds extended constantly along the direction of the axis of rotation,
but here we allow the e-folding decay depth of the cloud level wind,
$H(\theta)$, to vary with latitude. This enables extra degrees of
freedom in the possible structure of the winds compared to previous
studies such as \citet{Kaspi2010a,Kaspi2013a} and \citet{Liu2013}.
A latitudinal dependent decay depth can occur for several reasons,
such as the internal convection extent varying with latitude \citep[e.g.,][]{Aurnou2008},
ohmic dissipation being latitudinally varying \citep{Liu2008,Liu2010},
moist convection having different latitudinal behavior \citep{Lian2010},
or by different dynamics inside and outside the tangent cylinder surrounding
the metallic hydrogen envelope \citep[e.g.,][]{Heimpel2007,Gastine2013,Heimpel2015}.
The latitude dependent $H$ is defined as a summation over the first
20 Legendre polynomials
\begin{equation}
H\left(\theta\right)=\sum_{i=0}^{19}h_{i}P_{i}(\theta),\label{eq:depth of wind}
\end{equation}
where $h_{i}$ are the coefficients by which the shape of $H(\theta)$
is determined. Such formulation allows for a solution to be found
separately for different spatial scales of the winds and its resulting
gravity signals. Note that when setting $h_{i=1..19}=0$ the depth
of the winds is set to be constant with latitude. Since we expect
the dynamics to be in the regime of small Rossby numbers, the flow
to leading order is in geostrophic balance, and therefore thermal
wind balance must hold so that
\begin{equation}
\left(2\Omega\cdot\nabla\right)\left[\widetilde{\rho}\mathbf{u}\right]=\nabla\rho'\times\mathbf{g_{0}},\label{eq: thermal wind}
\end{equation}
where $\Omega$ is the planetary rotation rate, $\mathbf{u(r)}$ is
the full 3D velocity, $\mathbf{g_{0}}\left(r\right)$ is the mean
gravity vector\footnote{The gravity vector is calculated by integration of the static density
$\widetilde{\rho}$, and is therefore only a function of radius. \citet{Zhang2015}
suggest a correction to this equation by adding a term associated
with the non-radial component of the gravity vector due to dynamics.
However, as also shown by the same authors, for the values of decay
scale heights considered here and for gravity moments with $n>2$
such a term is small.} and $\rho'\left(r,\theta\right)$ is the dynamical density anomaly
\citep{Pedlosky1987,Kaspi2009}. Here the thermal wind balance is
written in a general form without making any assumptions on the depth
of the circulation. The mean static density $\tilde{\rho}(r)$ and
$\mathbf{g_{0}}\left(r\right)$ are calculated using the model of
\citet{Hubbard1999} (see also \citealp{Hubbard2014}). Note that,
in principle, the specific choice of these background fields affects
the dynamical density anomalies, however, we found that using $\tilde{\rho}(r)$
and $\mathbf{g_{0}}\left(r\right)$ from different sources hardly
affects the solution of the dynamical gravity field. Therefore we
consider $\tilde{\rho}(r)$ and $\mathbf{g_{0}}\left(r\right)$ as
known parameters and do not try to optimize them.

Integrating the zonal component of Eq. \ref{eq: thermal wind} latitudinally,
the dynamical density $\rho'\left(r,\theta\right)$ can be calculated,
and will depend only on the decay parameter $H\left(\theta\right)$
and a radially depending integration constant $\rho'_{0}\left(r\right)$,
which for small Rossby numbers is small compared to the solid-body
radial density profile, so that $\rho'_{0}\ll\tilde{\rho}$ \citep{Kaspi2013c}.
This integration constant, which physically represents a perturbation
to the horizontal-mean radial density profile due to dynamics, does
not contribute to the gravity field since it only depends on radius
while the Legendre polynomials are only functions of latitude with
a zero mean, and therefore in Eq.~\ref{eq: zonal harmonics-1} do
not contribute to the gravity moments. For this reason calculations
of the gravity signal using the thermal wind method as applied here
are limited to spherical geometry. Then, the dynamically induced gravity
moments due to the density anomaly $\rho'$ are
\begin{equation}
\Delta J_{n}=-\frac{1}{Ma^{n}}\intop_{0}^{a}r'^{n+2}dr'\intop_{0}^{2\pi}d\phi'\intop_{-1}^{1}P_{n}\left(\mu'\right)\rho'\left(r',\mu'\right)d\mu',\label{eq: dynamical zonal harmonics}
\end{equation}
using spherical coordinates so that $\phi$ is longitude and $\mu=\cos\theta$.
Note that unlike the dynamical gravity moments, the static gravity
moments are dominated by the oblate shape of the planet, and therefore
needs to be calculated by other methods \citep[e.g.,][]{Zharkov1978,Kong2012,Hubbard2012,Hubbard2014,Wisdom2016}.

In this study we will simulate the gravity moments, but it is important
to note that resulting gravity signal itself, which can be estimated
from radio tracking data, can then be calculated by taking the radial
and latitudinal derivatives of the gravity potential $V(r)=1-\sum_{n=2}^{\infty}\left(\frac{a}{r}\right)^{n}J_{n}P_{n}\left(\mu\right)$
to give the radial and latitudinal components of the anomalous gravity
perturbations due to wind given by 
\begin{equation}
\delta g_{r}=g_{0}\sum_{n=2}^{\infty}\left(n+1\right)\lambda^{n}\Delta J_{n}P_{n}\left(\mu\right),\label{eq:grav_anom_r_2}
\end{equation}
\begin{equation}
\delta g_{\theta}=g_{0}\sum_{n=2}^{\infty}\left(1-\mu^{2}\right)^{\frac{1}{2}}\lambda^{n}\Delta J_{n}\frac{dP_{n}}{d\mu},\label{eq:grav_anom_L_2}
\end{equation}
where $g_{0}$ is the mean surface gravity for the spherical planet,
$\lambda=a/(a+r_{p})$, and $r_{p}$ is the local distance in the
spacecraft's trajectory to the 1 bar surface. Note that for high moments
this signal increases rapidly as the spacecraft is close to periapse.

In summary, given the observed cloud-level winds, and assuming a penetration
depth of the winds and the dynamical balance between them and the
density structure, we can calculate the resulting gravity perturbation
on the planet surface. However, the problem we need to solve is the
inverse one: given the gravity measurements, what would be the $H(\theta)$
that would best explain them. For that, we develop the adjoint model
described in the next section.

\subsection{The adjoint model}

An efficient way to address the problem of determining the internal
structure of the wind field, given the observations of the gravity
moments $J_{n}$, a forward model such as described above, and the
observed cloud-level winds, is the adjoint method. This method allows
for an effective optimization of the model solution with respect to
a cost function and control variables \citep[e.g.,][]{Tacker1988,Tziperman1989,Tziperman1992,Wunsch2007,Mazloff2010}.
The adjoint method has been used extensively in geophysical fluid
dynamics problems on Earth, both in the ocean \citep[e.g.,][]{Marotzke1999,Galanti2003,Ferreira2005,Kalmikov2014},
and in the atmosphere \citep[e.g.,][]{Moore2011,blessing2014testing}.
It used for sensitivity studies as well as for optimization of parameters
and data assimilation.

The cost function is the physical quantity we wish to minimize. It
can be a measure of the deviation of the model solution from the observations,
or simply the model solution itself. The control variables can be
any parameter or model variable that has an effect on the cost function.
The adjoint model is then a backward run of the derivatives of the
cost function with respect to the model variables, linearized over
its solution from the forward integration, with the final solution
of the adjoint model being the sensitivity of the cost function with
respect to the control variables. This sensitivity can be studied
by itself, or be used to direct the model toward a solution that minimizes
the cost function. In this study we will use the adjoint mainly for
optimization, but also examine the adjoint sensitivities that enable
us to evaluate the sensitivity of our solutions to the wind velocities
at different depths.

The cost function is defined as the difference between the model calculated
moments and those measured, and the control variable is the decay
parameter $H\left(\theta\right)$. We define the cost function as
\begin{eqnarray}
{\cal J} & = & {\displaystyle \Delta J^{T}\cdot W}\cdot\Delta J+\epsilon\sum_{i=0}^{19}h_{i}^{2},\label{eq:cost_func}\\
\Delta{\cal J} & = & J^{c}-J^{o},
\end{eqnarray}
where $J^{c}$ is the $N$ size calculated model solution , $J^{o}$
are the observed gravity moments, and $W$ is a matrix of size $N\times N$
with weights given to each moment (diagonal terms) and covariance
between moments (off-diagonal terms). In general, the values of the
weights are set as the inverse of the observational error covariance
matrix \citep{Finocchiaro2010}, but given the conceptual nature of
this study we set for simplicity the weights to be $W_{ii}=4\times10^{16}$
and zero elsewhere, representing simulated uncertainties of $5\times10^{-9}$
(a value similar to the high moments, see more in section \ref{sec:Results}).
The second term in Eq.~\ref{eq:cost_func} which is set to be much
smaller than the first and controlled by the value of $\epsilon$,
acts as a constraint on the optimized solution demanding that the
values of $h_{i}$ be as small as possible as long as they do not
affect substantially the first term, thus reduce the effect of unphysical
initial guess on the final solution. For example, if a certain $h_{i}$
has a large value in the initial guess but has little effect on the
cost function, i.e., it has a small projection on the gravitational
moments, the second term acts to reduce its value. On the other hand,
if that $h_{i}$ has a significant projection on the gravitational
moments, and therefore on the cost function, it's value will be optimized.
This second term should only come into play once the cost function
has been substantially reduced, otherwise it will dominate the optimization
process and instead of minimizing the difference between the model
gravitational moments and the observed ones, the values of all $h_{i}$
will be reduced regardless of their contribution to the moments. It
is therefore important to keep the second term very small compared
to the initial value of the cost function (at least by 2 orders of
magnitude) to allow a physical optimization of the problem. With that,
the value of $\epsilon$ should also be not too small, otherwise it
will have no effect on the optimization process. For each case presented
here, the value of the parameter $\epsilon$ was set according to
the initial value of the cost function, to keep the necessary ratio.

Our goal is to minimize the cost function, i.e., bring the model solution
closer to the observed, and therefore we need to calculate its sensitivity
to changes in the decay parameter $H\left(\theta\right)$ . For simplicity
we start with a single $H$, so that the sensitivity is
\begin{equation}
\lambda\left(H\right)\equiv\frac{\partial{\cal J}\left(J_{n}^{c}\right)}{\partial H}\label{eq: adjoint parameter}
\end{equation}
with the modeled moments formulated as a series of operators
\begin{equation}
J_{n}^{c}=F_{1}(F_{2}(F_{3}(H))),\label{eq:model functions}
\end{equation}
where $F_{1}$ is the solution of Eq.~\ref{eq: dynamical zonal harmonics},
$F_{2}$ is the solution of Eq.~\ref{eq: thermal wind}, and $F_{3}$
the solution of the spherical structure of the wind Eq.~\ref{eq:zonal wind field},
which is a function of $H$.

Therefore the model Jacobian matrices for the three equations can
be written in the form \citep[e.g.,][]{Marotzke1999}
\begin{equation}
\lambda(H)\equiv\frac{\partial{\cal J}(H)}{\partial H}=\left(\frac{\partial{\cal J}}{\partial J_{n}^{c}}\right)\left(\frac{\partial J_{n}^{c}}{\partial\mathbf{\mathbf{\rho}}}\right)\left(\frac{\partial\mathbf{\mathbf{\rho}}}{\partial\mathbf{u}}\right)\left(\frac{\partial\mathbf{u}}{\partial H}\right).\label{eq:adjoint matrices}
\end{equation}
Taking the transpose results in matrix-vector multiplications and
we get
\begin{equation}
\lambda{}^{T}\equiv\left(\frac{\partial L}{\partial H}\right)^{T}=\left(\frac{\partial\mathbf{u}}{\partial H}\right)^{T}\left(\frac{\partial\mathbf{\rho}}{\partial\mathbf{u}}\right)^{T}\left(\frac{\partial J_{n}^{c}}{\partial\rho}\right)^{T}\left(\frac{\partial{\cal J}}{\partial J_{n}^{c}}\right)^{T},\label{eq:adjoint matrices-T}
\end{equation}
which is the adjoint model to be used.

The solution of the adjoint model $\lambda{}^{T}$ is the sensitivity
of the cost function to a perturbation in the control variable $H$.
Modifying the control variable iteratively according to the adjoint
solution will result in a minimization of the cost function. In the
case of a latitudinal dependent wind depth, where $H(\theta)$ is
a function of the coefficients $h_{i}$, the adjoint solution has
the form
\begin{eqnarray}
\lambda{}_{i}^{T}\equiv\left(\frac{\partial{\cal J}}{\partial h_{i}}\right)^{T} & = & \left(\frac{\partial\mathbf{H}}{\partial h_{i}}\right)^{T}\left(\frac{\partial\mathbf{u}}{\partial H}\right)^{T}\left(\frac{\partial\mathbf{\rho}}{\partial\mathbf{u}}\right)^{T}\nonumber \\
 & \cdot & \left(\frac{\partial\Delta J_{n}^{c}}{\partial\rho}\right)^{T}\left(\frac{\partial{\cal J}}{\partial\Delta J_{n}^{c}}\right)^{T}+2\epsilon h_{i},\label{eq:adjoint matrices-T-hc}
\end{eqnarray}
where $\lambda{}_{i}^{T}$ is the adjoint sensitivity with respect
to the $i^{th}$ coefficient of Eq.~\ref{eq:depth of wind}. Alternatively,
the adjoint sensitivities could be formulated using Lagrange multipliers
\citep[e.g.,][]{Tacker1988,Tziperman1989}. A detailed example on
how the adjoint model is derived using the Lagrange multipliers is
given in Appendix~\ref{Appendix-A}.

The effectiveness of the adjoint method comes from its ability to
provide the sensitivity to all control variables (in our case, $h_{i}$)
in a single run of the forward and backward models. Having the adjoint
solution we can now proceed to construct the optimization of the model
solution.

\subsection{Optimization procedure	\label{sub:Optimization-procedure	}}

\begin{figure}[h]
\begin{centering}
\includegraphics[width=1\columnwidth]{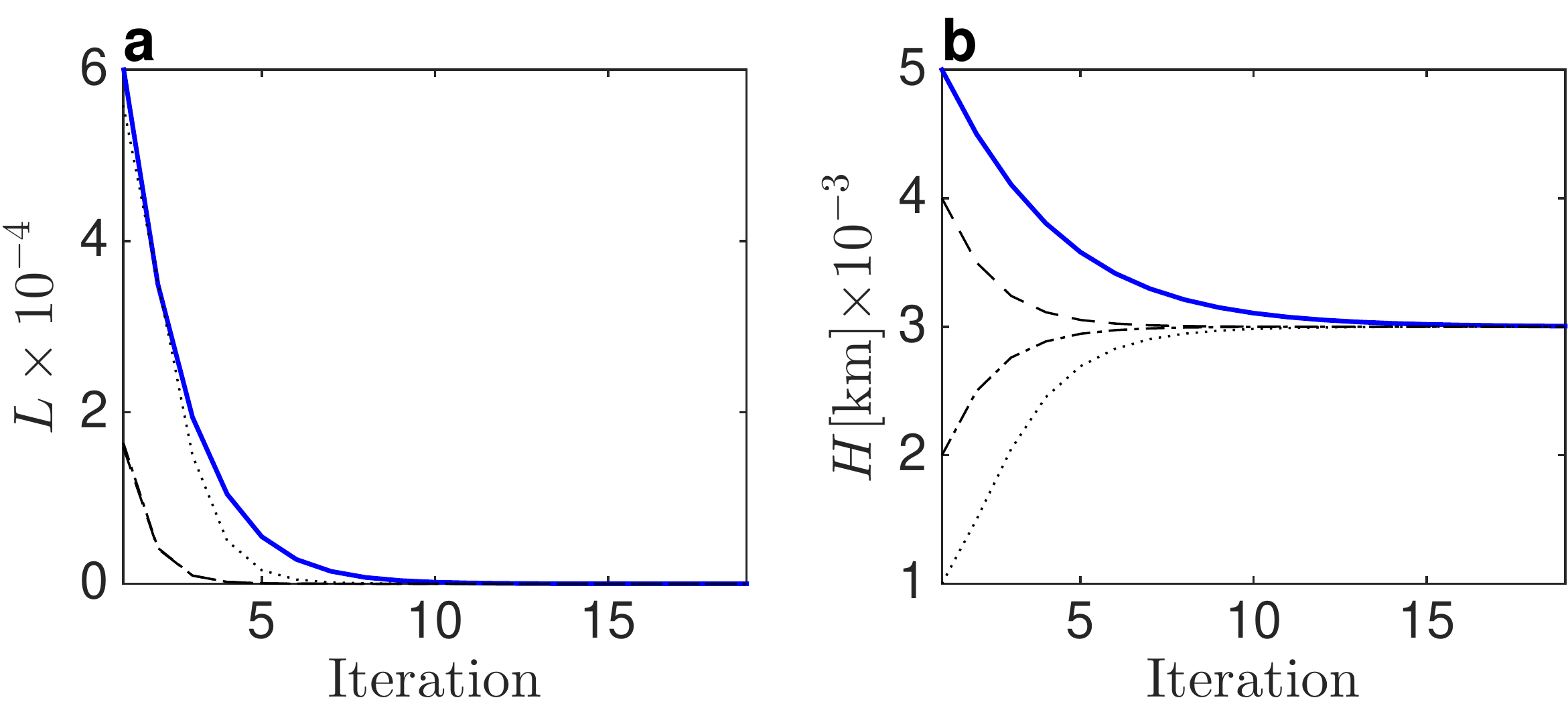}
\par\end{centering}

\caption{An example of the adjoint optimization. (a) the reduction of the cost
function with iterations, and (b) the change in the penetration depth
as function of iterations. The solid line is for the optimization
starting from an initial guess of $H=5000\,\textup{km}$, while the
other lines show the same optimization but with the initial guess
being $H=4000\,\textup{km}$, $H=2000\,\textup{km}$ and $H=1000\,\textup{km}$.
In all cases, the same solution is reached. Note that due to the nonlinear
nature of the problem the convergence rate is different for the cases
of initial guess of $H=5000\textup{\,km}$ and $H=1000\,\textup{km}$,
even though they begin from the same distance from the solution. Also
note that the dashed and dashed-dotted lines in (a) overlap.\label{fig:Example-of-adjoint}}
\end{figure}

\begin{figure}[h]
\begin{centering}
\includegraphics[width=1\columnwidth]{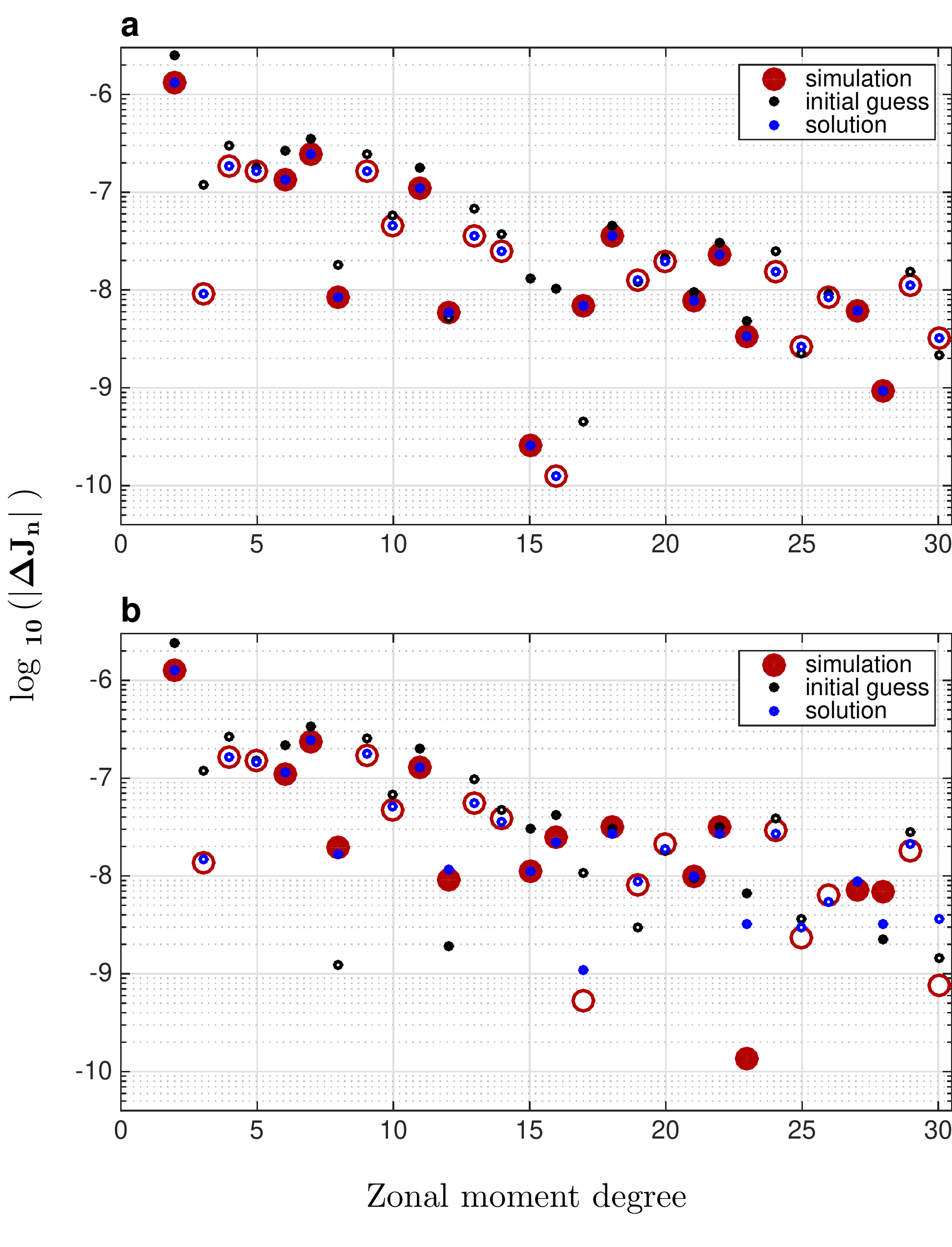}
\par\end{centering}

\caption{The Jupiter gravitational moments for (a) a case with no errors in
the simulated moments, and (b) a case with random errors. Shown are
the simulation (red), initial guess (black), and solution (blue).
Filled (open) circles denote positive (negative) values. \label{fig:Jupiter-Hsingle-moments}}
\end{figure}

Once the gradient of the cost function $\lambda$ is obtained, the
control variables (either a single depth $H$, or coefficients $h_{i}$)
are modified so that in the next iteration the cost function will
have a lower value. In the case of a single $H$ there is only one
option to change the control variable - in the direction opposite
to the value of the adjoint solution. In the case of optimizing $h_{i}$,
moving directly (steepest descend) is not efficient, therefore a conjugate
gradient method is applied so that the direction of modifying the
control variables is the optimal one \citep{Herstenes1980}. The extent
of the change is also controlled using a line search \citep{Herstenes1980},
so that the change in the control variable does not cause the cost
function to move beyond the global minimum. The global minimum is
defined to be reached when each of the gravitational moments is as
close to the value of the observed one as the size of the uncertainty
assigned to it. Therefore, after each iteration we check the value
of each element in $\Delta J$ to see whether it is small compared
to the observational uncertainty (in our case, $5\times10^{-9}$),
and if all the calculated moments are within this uncertainty the
optimization is complete. At the final iteration, the Hessian matrix
$C$ (second derivative of the cost function) is calculated in order
to estimate the uncertainties associated with each control variable
\citep{Tziperman1989}. Inverting the Hessian matrix $C$, we get
the error covariance matrix $G$. Finally, the cross correlated uncertainties
$G_{ij}$ are projected into the physical space of $H$, and used
as formal bounds on the solution. Note that if the control variable
is a single depth $H$, the size of the Hessian matrix is $1\times1$.
The adjoint optimization was tested with various wind depths, and
is found to be able to reach a solution within 10 to 60 iterations
(see example in Figs.~\ref{fig:Example-of-adjoint} and \ref{fig:Jupiter-Hsingle-moments}).
In addition, an important criteria for the robustness of the solution
is whether the minimum reached is global, i.e., would the same solution
be reached if starting from different initial guesses. In order to
test this we tested each of the experiments presented in this study,
starting from different initial guesses and checking if the same solution
is reached. An example of such a test is shown in Fig.~\ref{fig:Example-of-adjoint},
where the different curves show the optimization process for different
initial guesses. We found that in all experiments discussed in section~\ref{sec:Results}
the adjoint optimization is insensitive to the choice of the initial
guess - the same solution is being reached regardless of the initial
guess, only the number of iterations needed might change. Therefore,
we conclude that the global minimum of the problem is indeed being
reached, and the method is valid for the problems presented.

\section{Results}
\label{sec:Results}

We now show how the adjoint method is used to produce a wind structure,
given the surface wind velocities and a set of gravity moments. In
all cases presented below the following approach was taken: first,
the forward model was run with a chosen depth of winds and the gravitational
moments were calculated. These moments are then defined as the \textbf{\textit{simulation}}.
This solution is used to mimic the upcoming Juno or Cassini observations.
In order to allow for observational errors, a uniformly distributed
error with a magnitude of $5\times10^{-9}$ is added to the gravity
moments. This value corresponds to the average error measurement we
expect for lowest order moments \citep{Finocchiaro2010}. Second,
the adjoint optimization is used to find the \textbf{\textit{solution}}
closest to the simulated moments, starting from an \textbf{\textit{initial}}\textbf{
}\textbf{\textit{guess}} and then searching for the solution using
the optimization procedure.

\subsection{Inversion of the gravity signal}

We start with a Jupiter case, where first for simplicity the observed
winds of Jupiter are set to penetrate to the same depth in all latitudes.
Therefore the adjoint optimization is trying to minimize the cost
function with respect to a single depth $H$. The simulation is done
with $H=3000\,\textrm{km}$, and the initial guess was set to $H=5000\,\textup{km}$.
The progression of the adjoint optimization is shown in Fig.~\ref{fig:Example-of-adjoint}
(solid lines). It can be seen that the within 5 iterations the cost
function value is reduced by an order of magnitude, and the calculated
wind depth is getting closer to the simulated one. The gravitational
moments for the simulation, initial guess, and solution are shown
in Fig.~\ref{fig:Jupiter-Hsingle-moments} for a case without errors
in the simulated moments (panel a) and for a case with errors (panel
b). While in the former case the moments reach a perfect fit to the
simulation, in the latter case the lower moments are fitted well and
the higher moments are less so, since the errors applied to the simulation
have a relatively larger effect on the higher moments which are smaller.

\begin{figure}[h]
\begin{centering}
\includegraphics[width=1\columnwidth]{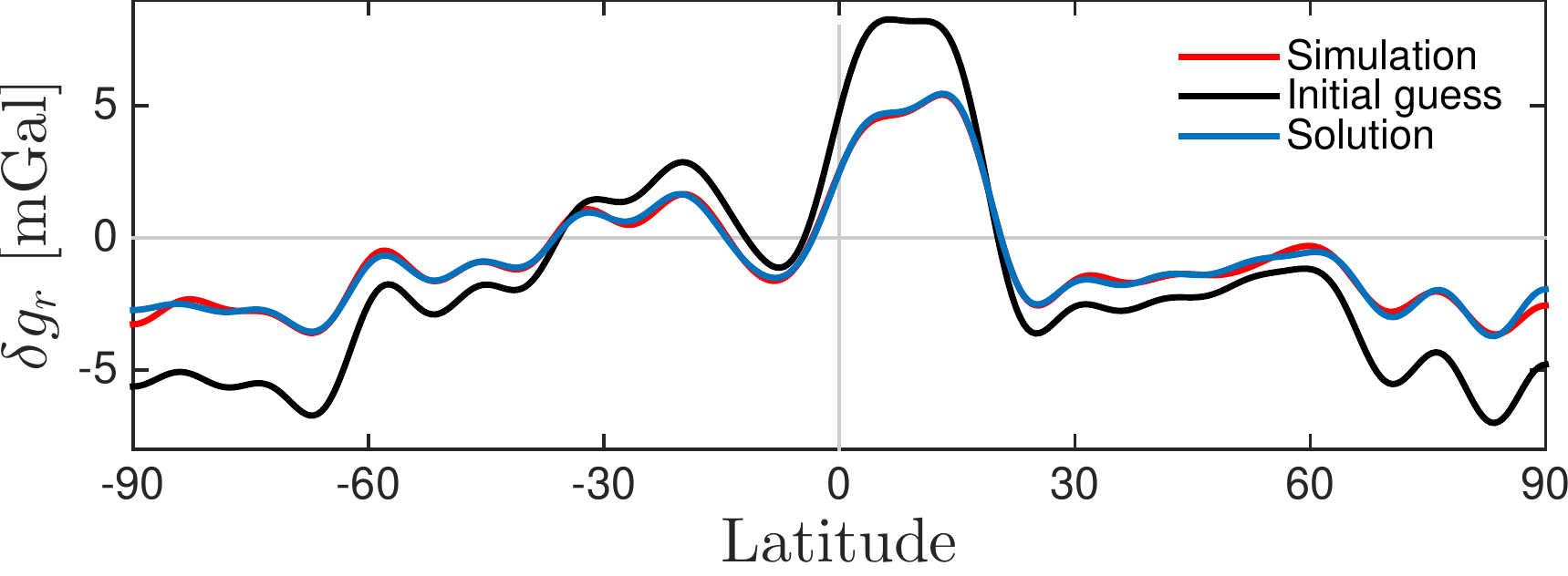}
\par\end{centering}

\caption{A case with simulated $H=3000\,\textrm{km}$ and the initial guess
$H=5000\,\textup{km}$. Shown are the radial component of the gravitational
anomalies (in mGals) as function of latitude, for the simulation (red),
initial guess (black), and solution (blue). \label{fig:Jupiter-Hsingle-gravitational-anomalies}}
\end{figure}

The gravitational moments can be transformed into the actual latitudinal
dependent gravity anomalies (Eq.~\ref{eq:grav_anom_r_2}). The radial
component of these anomalies, calculated at the planet's 1~bar level,
are shown in Fig.~\ref{fig:Jupiter-Hsingle-gravitational-anomalies},
for the adjoint solution, together with the simulation and initial
guess. Note that as a result of the complex surface wind structure
(Fig.~\ref{fig:winds}), even with a single $H$ the resulting gravity
field has a pronounced asymmetry between the Northern and Southern
hemispheres. While the initial guess differs considerably from the
simulation, the solution matches well the simulation. The difference
between the simulation and the solution is a result of the errors
we apply to the simulation. In a case with no random errors the solution
is identical to the simulation (not shown). Given the good agreement
between the adjoint solution and the simulation, it is clear that
most of the gravity signal is contained in the lower moments. Note
that the differences in the moments (Fig.~\ref{fig:Jupiter-Hsingle-moments})
have only minor effect on the actual gravity field; therefore, for
the next more complicated cases, we show the gravity field and not
the gravitational moments.

Next, the winds penetration depth is allowed to vary with latitude.
In this case, the adjoint optimization is trying to minimize the cost
function with respect to the set of coefficients $h_{i}$. The simulation
was done with an $H(\theta)$ that is deeper at the equator and shallower
toward the poles, representing possibly deeper dynamics outside of
the tangent cylinder \citep{Aurnou2007}, where drag might be playing
a lesser role \citep{Liu2010}. The simulation $H\left(\theta\right)$
has an asymmetry between the hemispheres, and the initial guess was
chosen to reflect a complex dependence on latitude (Fig.~\ref{fig:Jupiter-Hlat-gravitational-anomalies}b).
The adjoint solution over most latitudes (see below) is in good agreement
with the simulation, in both the gravity anomalies (Fig. \ref{fig:Jupiter-Hlat-gravitational-anomalies}a)
and the depth of the winds (Fig. \ref{fig:Jupiter-Hlat-gravitational-anomalies}b). 

\begin{figure}
\begin{centering}
\includegraphics[width=1\columnwidth]{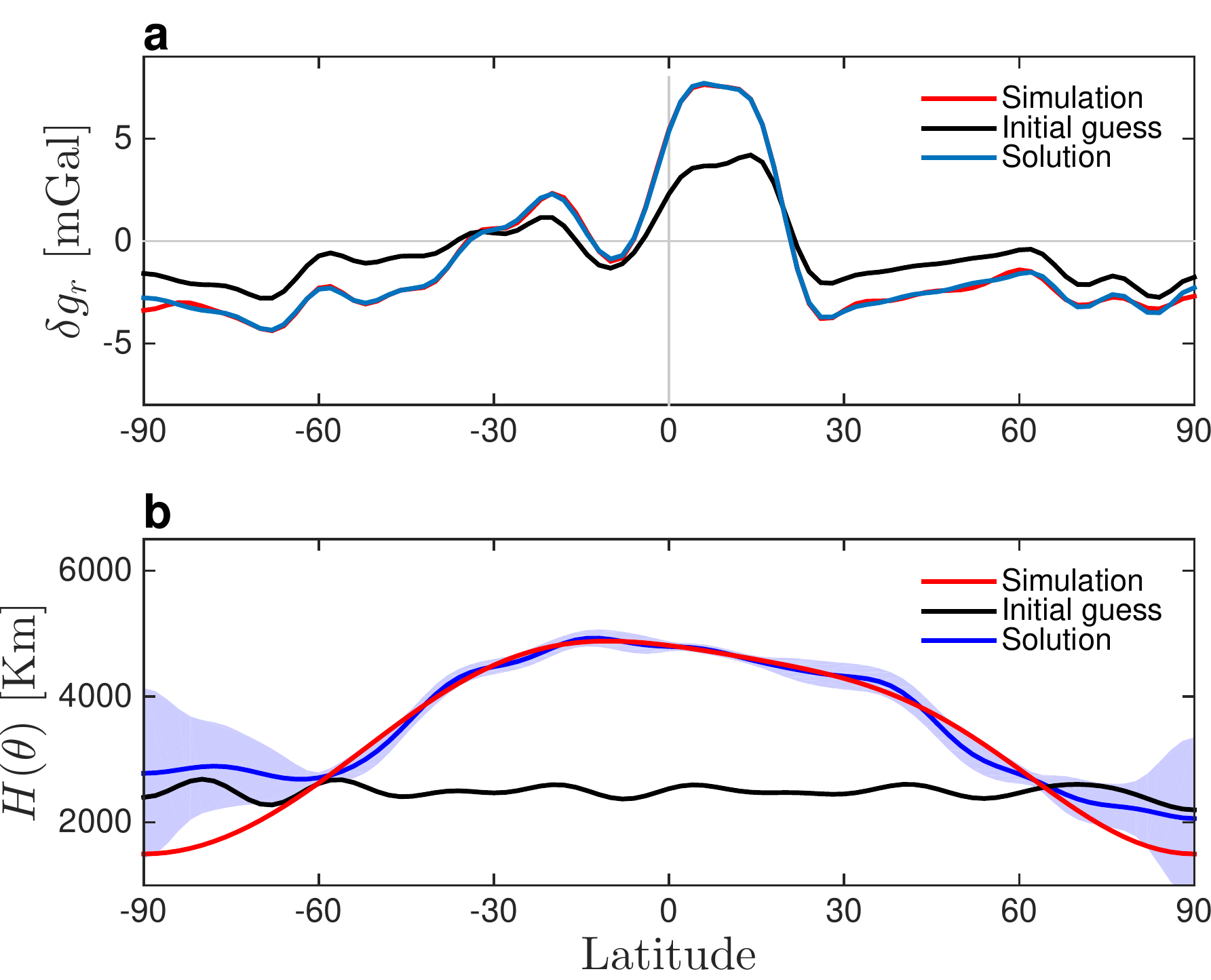}
\par\end{centering}

\caption{A case with latitudinal depended decay depth $H(\theta)$. (a) The
gravitational anomalies (in mGals) as function of latitude, and (b)
the depth of the winds. Shown are the simulation (red), initial guess
(black), and solution (blue). For the solution of the wind depth,
also shown are the associated uncertainties (blue shading). \label{fig:Jupiter-Hlat-gravitational-anomalies}}
\end{figure}

\begin{figure}
\begin{centering}
\includegraphics[width=1\columnwidth]{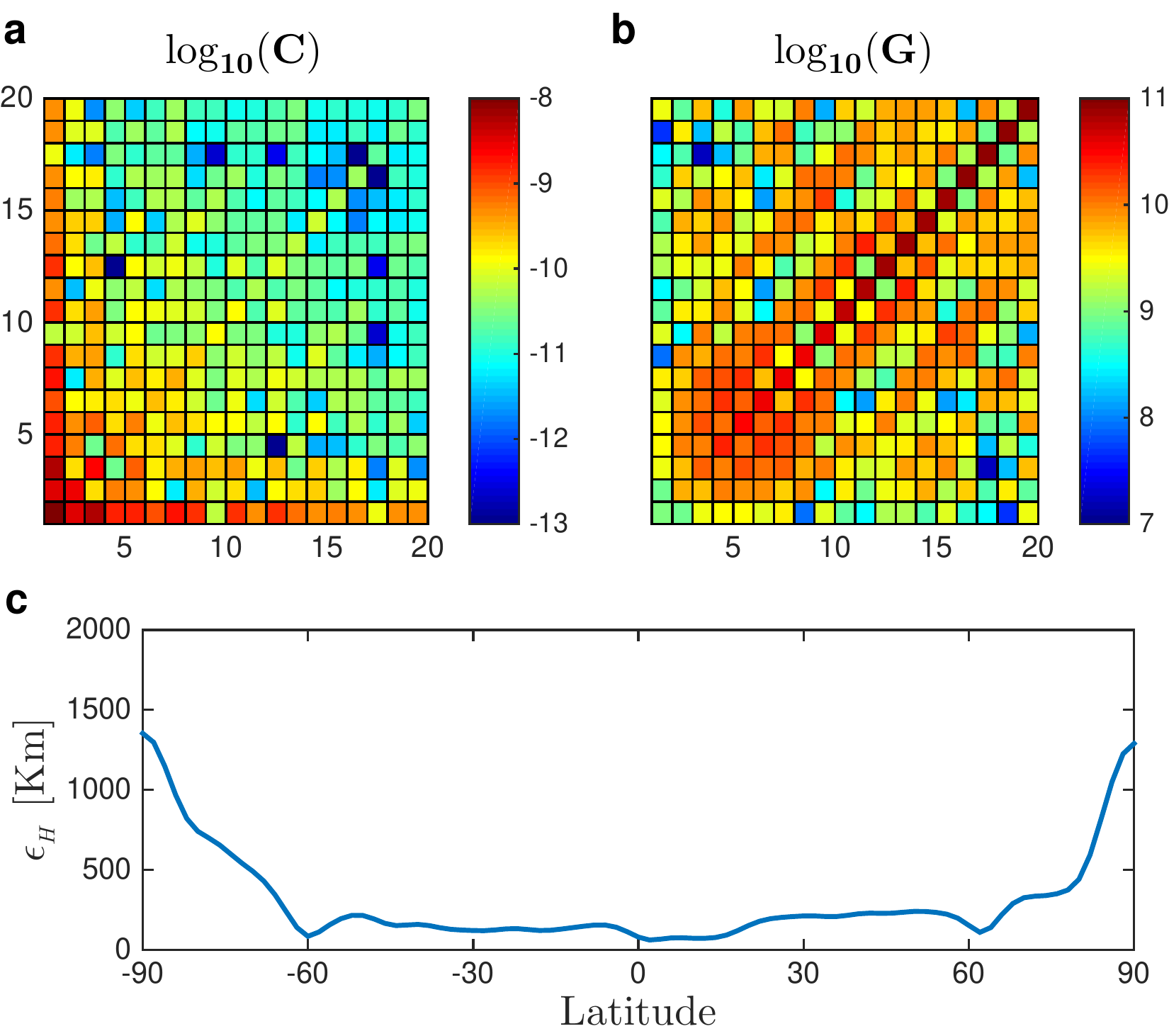}
\par\end{centering}

\caption{(a) Hessian matrix associated with the adjoint solution. (b) The error
covariance matrix. (c) The uncertainties in the solution for the depth
of the winds, as calculated from the diagonal of the error covariance.
\label{fig:Jupiter-Hlat-uncertainties}}
\end{figure}

The adjoint model also gives us the ability to rigorously estimate
the uncertainty associated with the solution. This is done by calculating
the Hessian matrix and from that, the error covariance matrix (Fig.~\ref{fig:Jupiter-Hlat-uncertainties}a
and Fig.~\ref{fig:Jupiter-Hlat-uncertainties}b, respectively). The
Hessian matrix gives an estimate to the sensitivity of the cost function
to cross perturbations in the control variables; the higher the value,
the better is our ability to determine the values of these parameters.
Inverting the Hessian matrix reveals the covariance of the uncertainties
associated with each control variable (Fig.~\ref{fig:Jupiter-Hlat-uncertainties}b).
The highest uncertainties are in the diagonal terms of the higher
coefficients, meaning that the coefficients are not affecting much
each other, and that most of the uncertainties are in the spatially
highly variable features (higher moments). In order to further verify
this, the covariance matrix can be used to calculate the uncertainty
in $H$ at each latitude using
\begin{equation}
\epsilon_{H}(\theta)=\sqrt{\sum P_{i}(\theta)P_{j}(\theta)G_{ij}}\label{eq:H uncertainties}
\end{equation}
where $G_{ij}$ are the cross correlation between the coefficients
$h_{i}$ and $h_{j}$, and $P_{i}(\theta)$ is the Legendre polynomial
at latitude $\theta$. The uncertainty in $H$ (Fig.~\ref{fig:Jupiter-Hlat-uncertainties}c),
is found to be largest near the poles since this is where the winds
are weakest. This uncertainty is plotted also in Fig.~\ref{fig:Jupiter-Hlat-uncertainties}b
to illustrate the usefulness of the solution for $H$. It is clear
that where the uncertainty in $H$ is small, the solution matches
well the simulation (low latitudes), and where the uncertainty is
large the solution deviates considerably from the simulation (high
latitudes). This has implications for the expected usefulness of the
upcoming gravitational measurements by Juno and Cassini in determining
the depth of the winds on both Jupiter and Saturn. Even in a case
where the uncertainties of the measured gravity field is very small,
due to the weak wind close to the poles it will not be possible to
determine with certainty the depth of the flow in the polar regions.
Moreover, the planned periapses of the orbit of both Juno and Cassini
are at low latitudes, meaning that the sensitivity to higher latitude
induced signals will be smaller \citep{Finocchiaro2010,Finocchiaro2013}.

\begin{figure}
\begin{centering}
\includegraphics[width=1\columnwidth]{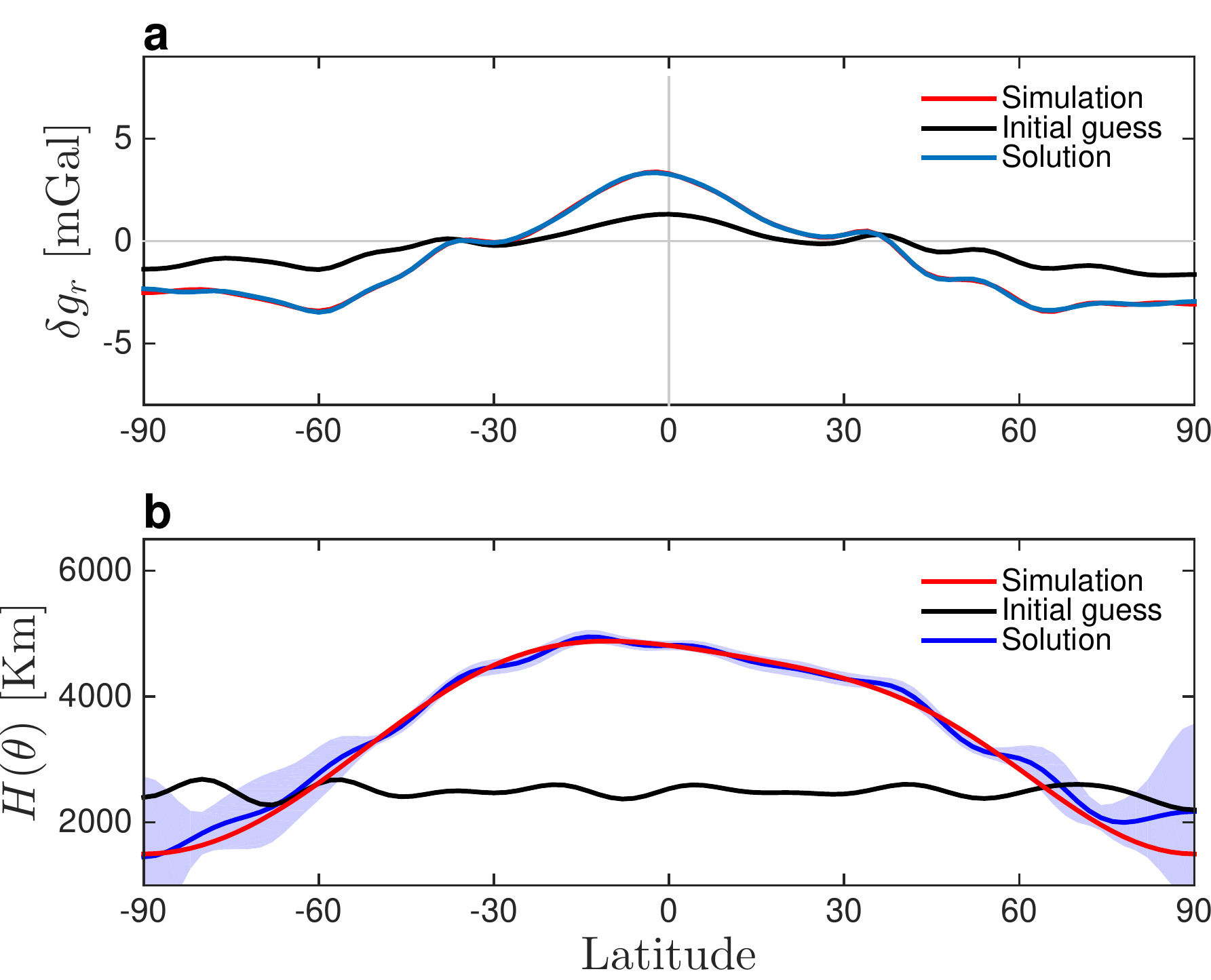}
\par\end{centering}

\caption{(a) The Saturn gravitational anomalies (in mGals) as function of latitude,
and (b) the depth of the winds. Shown are the simulation (red), initial
guess (black), and solution (blue). For the solution of the wind depth,
also shown are the associated uncertainties (blue shadow). \label{fig:Saturn-Hlat-gravitational-anomalies}}
\end{figure}
Next, we repeat the experiment for Saturn. The main difference between
the planets, aside from the different physical parameters, is in the
surface wind pattern \citep{Sanchez-Lavega2000,SanchezLavega2007,Garcia-Melendo2011}.
While on Jupiter the winds vary considerably with latitude, on Saturn
the much stronger winds have a simpler latitudinal pattern (Fig.~\ref{fig:winds}).
As in the Jupiter case, the simulation was done with $H$ that is
deeper at the equator and shallower toward the poles, with an asymmetry
between the hemispheres; the initial guess was chosen to reflect a
complex dependence on latitude (Fig.~\ref{fig:Saturn-Hlat-gravitational-anomalies}b).
Since the shape of the surface winds are much different in the Saturn
case, the resulting gravity anomalies are also very different (Fig.~\ref{fig:Saturn-Hlat-gravitational-anomalies}a).
Still, similarly to the Jupiter case, the solution matches closely
the simulation in the low latitudes and less so closer to the poles.
This characteristic is evident in the associated uncertainties, shown
in Fig.~\ref{fig:Saturn-Hlat-gravitational-anomalies}b as the shaded
area around the solution.

\subsection{Changing the physical assumptions}

So far we used the same physical assumptions in both the model used
for calculating the 'simulation' and the model used to find the 'solution'.
However, a valid question is how well would the adjoint optimization
work when the model used to find the solution is different from the
one used for generating the simulation - in the reality we should
expect that any model used to interpret Juno observations will lack
some of the physics embedded in the observations.

\begin{figure}[h]
\begin{centering}
\includegraphics[width=1\columnwidth]{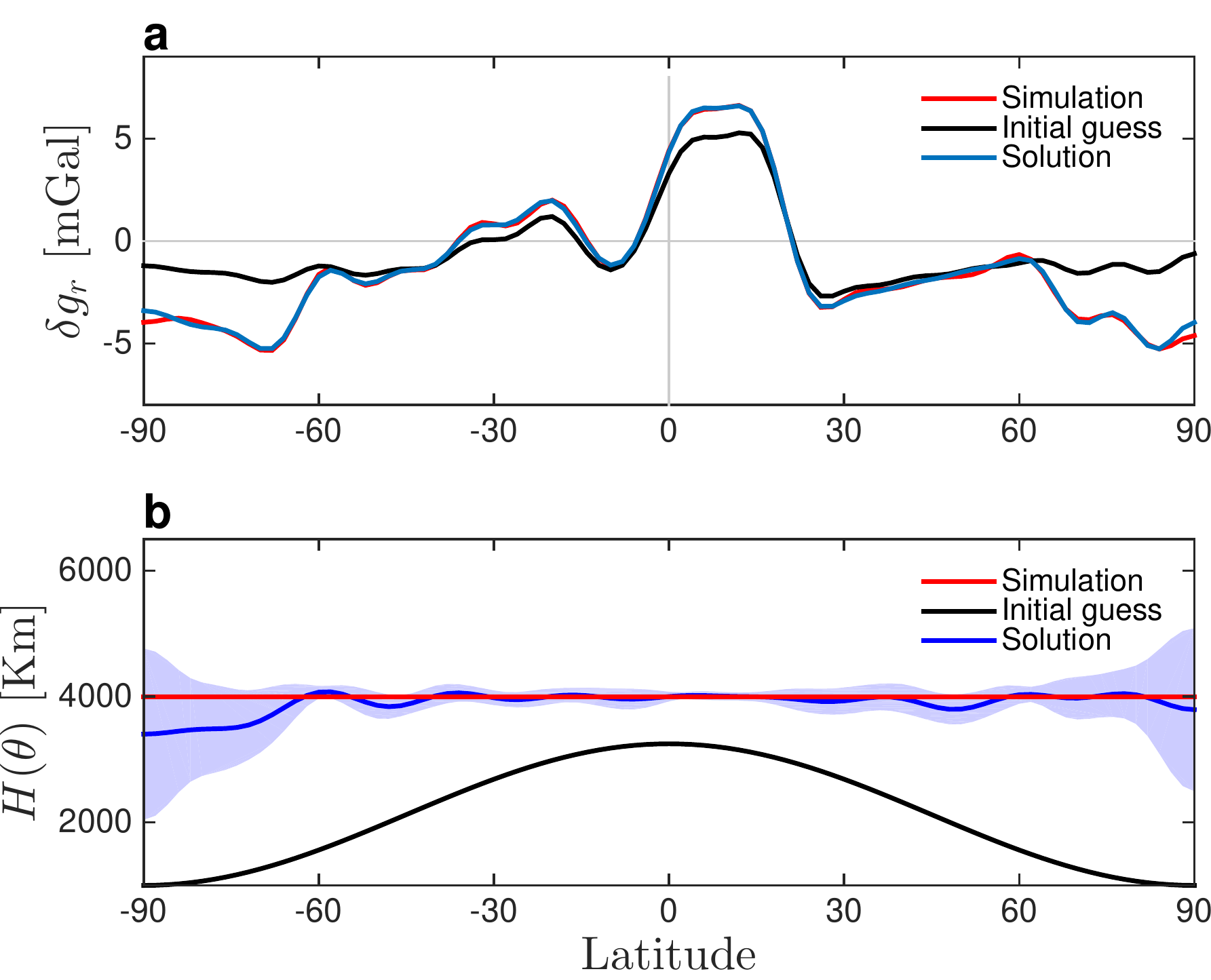}
\par\end{centering}

\caption{A case where the simulated wind depth is constant with latitude and
the solution varies with latitude. (a) Jupiter gravitational anomalies
(in mGals) as function of latitude, and (b) the depth of the winds.
Shown are the simulation (red), initial guess (black), and solution
(blue). For the solution of the wind depth, also shown are the associated
uncertainties (blue shadow). \label{fig:Jupiter-H_Hlat-gravitational-anomalies}}
\end{figure}

In order to get insight on the matter, we examine two cases in which
the model used for optimization differs from that used for the simulation.
First, we set the 'simulation' with the depth of the wind being constant
with latitude, and ask the optimization to look for a depth that varies
with latitude (Fig.~\ref{fig:Jupiter-H_Hlat-gravitational-anomalies}).
The simulation was done with $H=4000\,\textup{km}$, the initial guess
was set with depth of winds that vary between $3000\,\textup{km}$
near the equator and $1000\,\textup{km}$ near the poles. The optimized
solution follows closely the simulated depth, aside from the polar
regions where it deviate by about $1000\,\textup{km}$ (as expected
from its uncertainties). This experiment challenges the optimization
more than the previous ones, but is still within the framework of
the physical model used for both simulation and optimization, since
we can view the simulation as done with latitude varying depth set
with Eq.~\ref{eq:depth of wind} with $h_{0}=4000\,{\rm km}$ and
$h_{1-19}=0$.

\begin{figure}[t]
\begin{centering}
\includegraphics[width=1\columnwidth]{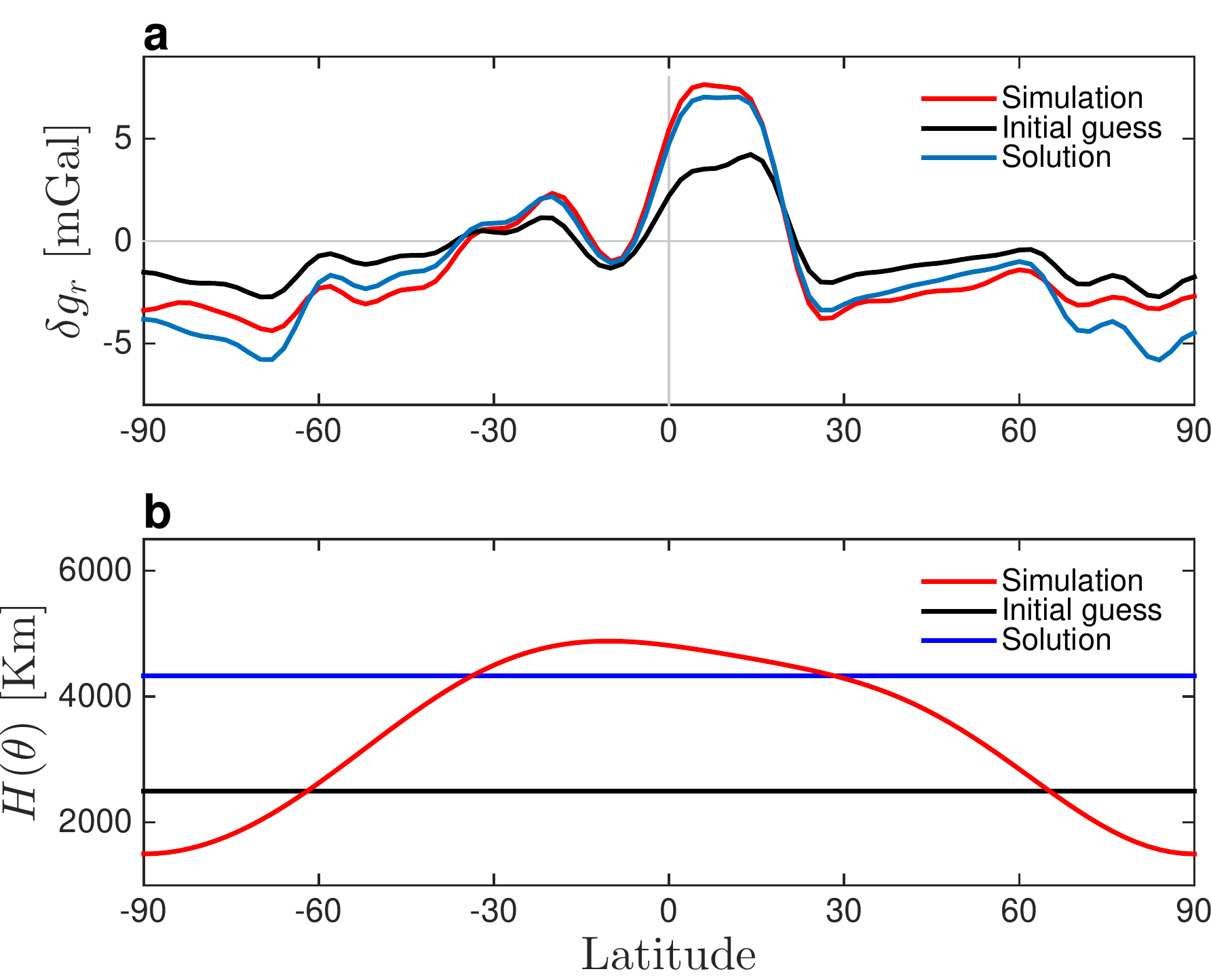}
\par\end{centering}

\caption{A case where the simulated wind depth varies with latitude but the
solution has a constant depth. (a) Jupiter gravitational anomalies
(in mGals) as function of latitude, and (b) the depth of the winds.
Shown are the simulation (red), initial guess (black), and solution
(blue). \label{fig:Jupiter-H_Hlat1-gravitational-anomalies}}
\end{figure}

A more challenging setup can be done by generating the simulation
with a latitude varying depth and looking for a solution in which
the depth of the wind is assumed constant with latitude. In such a
case we can expect that the simulated solution could not be reached,
since the physical model used in the optimization lacks some of the
physics used in the simulation. We set the experiment with the simulation
based on the same depth distribution as in the previous section, and
set the initial guess to be $H=2000\,{\rm km}$ (Fig.~\ref{fig:Jupiter-H_Hlat1-gravitational-anomalies}).
As expected, it can be seen that the solution $H=4327\,\textup{km}$
does not match the simulated one, yet it is in the proximity of the
wind depth in the equatorial region. Moreover, looking at the actual
gravity field (Fig.~\ref{fig:Jupiter-H_Hlat1-gravitational-anomalies}a),
the solution captures most of the signal contained in the simulation,
especially in the low and mid-latitudes.

The two experiments discussed here show that the adjoint method can
deal with cases in which the physical assumptions regarding the depth
of the wind used in the optimization differ from those used in the
simulation. In the next section we discuss the sensitivity of the
solution to deep flow patterns, an additional complication.

\subsection{Sensitivity to deep wind pattens}

Aside from the question of how deep the cloud-level winds penetrate,
it might also be the case that a different structure of flow exists
in the interior, that does not have any signature at the observed
cloud-level wind. As deeper levels have more mass, it is possible
that the measured gravity signal will come from these levels \citep{Galanti2015DPS}.
As long as this flow structure will be large scale it will also likely
be geostrophic and therefore in thermal wind balance, with an associated
density modulation that will affect the gravity field. Using the adjoint
method we can get an estimate on the sensitivity of the cost function
to perturbations in the 2-dimensional wind field, regardless of the
surface winds. In fact this can be also done in 3D, using the Tesseral
moments \citep{Parisi2016}, but as the aim of this study is to present
this method, we keep here the analysis simple with just the 2D (zonally
averaged) wind fields. To perform this estimate, the forward model
was run with a wind depth of $5000$~km (this choice does not affect
the solution, aside from some minor nonlinear contributions). Defining
the cost function to be the sum of the square of the gravitational
moments
\begin{equation}
{\cal J}=\sum_{n}W_{n}(J_{n}^{c})^{2},\label{eq:cost function-2}
\end{equation}

the adjoint model is integrated to produce the adjoint sensitivities,
which are the sensitivity of the cost function to any of the model
prognostic variables. In our case they include the gravitational moments,
the density perturbations, the wind structure, and the depth parameter
(see Eq.~\ref{eq:adjoint matrices-T}). The value of the sensitivities
is the change in the cost function expected when perturbing the variable
with a unit change.

\begin{figure}[t]
\begin{centering}
\includegraphics[width=1\columnwidth]
{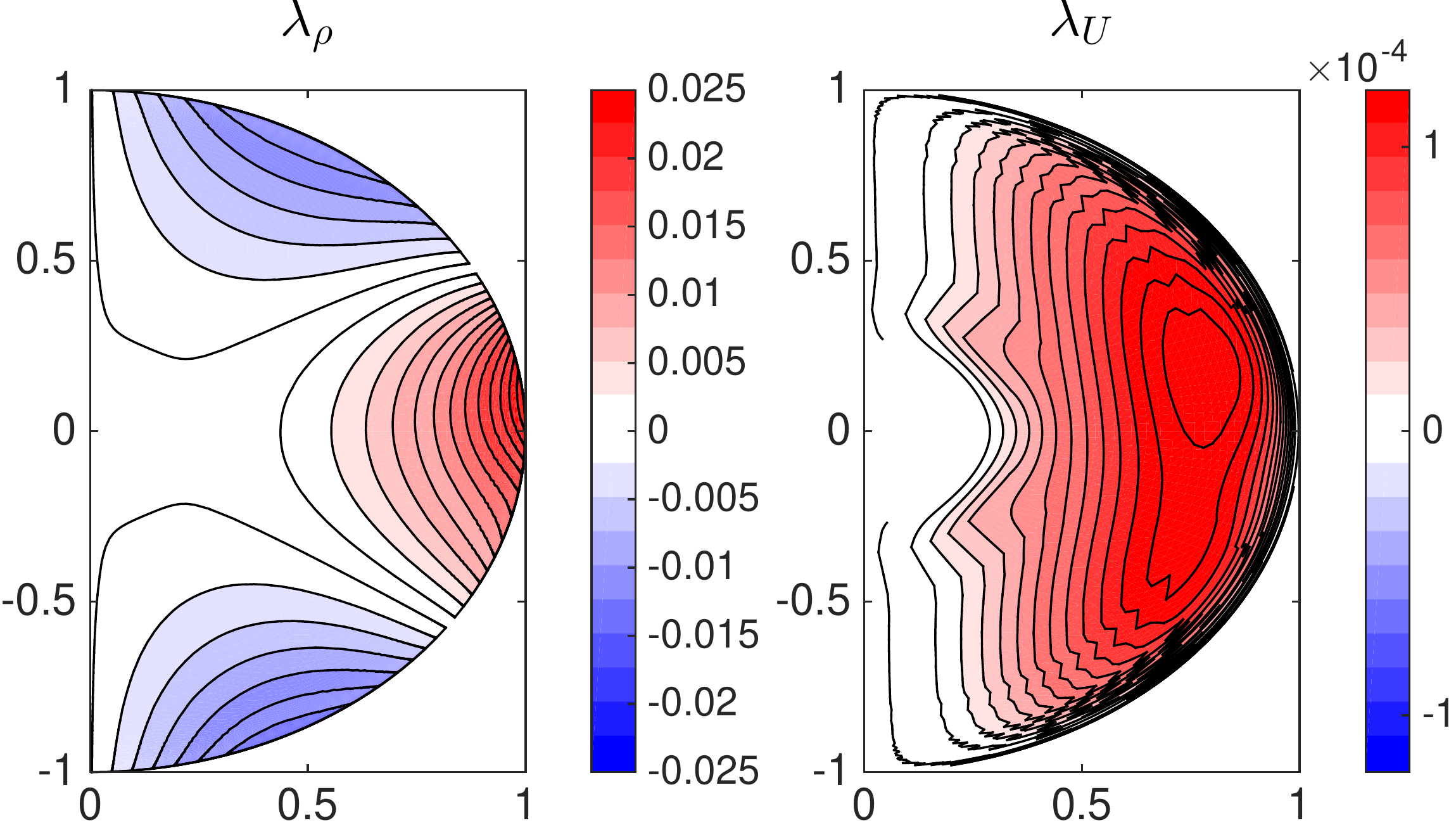}
\par\end{centering}

\caption{Sensitivity of the cost function to perturbations in $\rho$ (left)
and $U$ (right). \label{fig:Jupiter-rho-U-sensitivities}}
\end{figure}

We then save the adjoint variables of both the density and wind,
\begin{eqnarray}
\lambda_{\rho}(z,\theta) & \equiv & \left(\frac{\partial J^{c}}{\partial\rho}\right)^{T}\left(\frac{\partial{\cal J}}{\partial\Delta J^{c}}\right)^{T},\,\nonumber \\
\lambda_{U}(z,\theta) & \equiv & \left(\frac{\partial\mathbf{\rho}}{\partial\mathbf{u}}\right)^{T}\left(\frac{\partial J^{c}}{\partial\rho}\right)^{T}\left(\frac{\partial{\cal J}}{\partial J^{c}}\right)^{T},\label{eq:adjoint of U}
\end{eqnarray}
that are a function of latitude and depth. The adjoint solution for
the sensitivity to the density and wind is specific to the model's
numerical structure in general, and to the grid structure in particular.
Given that the grid in the model is not regular, i.e., the size of
the grid box changes with depth, the adjoint solution needs to be
normalized by the size of the grid box in order to show the physical
sensitivity. Figure~\ref{fig:Jupiter-rho-U-sensitivities} shows
the normalized sensitivities to density perturbations and wind perturbations.
While the sensitivities to the density mostly bears the shape of $J_{2}$,
and are highest close to the surface (due to the strong dependence
of the gravitational moments on the distance from the surface), the
sensitivities to the zonal velocity show a different structure. There
exist a single pattern of positive sensitivities, located close to
the equator, with the maximum around a depth of $10,000$~km from
the surface. The sensitivities decay gradually toward high latitudes.
This is due to the nature of the thermal wind balance, in which the
vertical gradient of the density is a function of the latitudinal
gradient of the wind, so that a wind perturbation at a certain depth
will imply a density perturbation from that depth to the surface.
This implies that the highest sensitivities to deep winds, if they
exist, will be to winds at the range of $\sim10,000\,{\rm km}$ below
the cloud-level and limited to low latitudes. This issue, of the possibility
that deep flows exists separately from the surface wind and their
affect on the gravity field, needs to be further examined in future
studies.

\section{Discussion and conclusion}
 \label{sec:Discussion-and-Conclusion}

Modern observations of the gas giants since the 1970s have allowed
studying these planets in great detail, particularly regarding processes
at their cloud-level. Yet, much of the processes controlling the levels
below have remained unknown mainly due to lack of observational data.
This will likely change in the coming few years as the upcoming Juno
mission and Cassini proximal orbits bring the possibility of investigating
in detail the sub-cloud levels with several different instruments.
Particularly, Radio measurements will provide high precision gravity
soundings, i.e., data that can be used to estimate the depth of the
observed surface flows on these planets. All models to date, relating
the winds to the gravity field, have been in the forward direction,
thus allowing only a calculation of the gravity field from a given
wind model. Here, we propose a method to solve the inverse problem
of deriving the depth of the winds from the gravity data. We use an
adjoint based inverse dynamical model to relate the expected measurable
gravity field, to perturbations of the density and wind fields, and
therefore to the observed cloud-level winds. In order to invert the
gravity field to be measured by Juno and Cassini into the circulation,
an adjoint model is constructed for the dynamical model, thus allowing
backward integration of the thermal wind model. 

The thermal wind method allows perhaps the simplest relation between
the flow velocity and the dynamically balanced density gradients,
which can be then related to the gravity field. Therefore, we have
applied the adjoint method to the thermal wind model; however, the
methodology presented here is not specific to this model and could
be used to study and optimize any type of model, ranging from simple
conceptual models to complex general circulation models \citep[e.g.,][]{Marotzke1999,Galanti2003,Mazloff2010}.
In any such model the adjoint method will allow backward calculation
of the flow field that best matches the measured gravity field. Models
with more complex physics will allow the inclusion of processes not
taken into account here such as magnetic effects, internal convection
etc. Nonetheless, as long as the large scale motion on these planets
is controlled to leading order by the rotation of the planet, even
models containing more physical processes would be to leading order
in thermal wind balance. Therefore, this analysis captures the leading
order dynamical balance between the gravity and the winds. More sophisticated
wind structures with more complex depth and latitudinal dependencies,
and weaker coupling to the cloud-level wind can also be considered.

This tool proves to be useful for various scenarios, simulating cases
in which the depth of the wind is constant, or varies with latitude.
We show that it is possible to use the gravity measurements to derive
the depth of the winds, both on Jupiter and Saturn, taking into account
also measurement errors. We find that due to the winds on both planets
being much stronger in the equatorial regions, the model solutions
are better determined in the low to mid-latitudes, while the depth
of the winds close to the poles cannot be determined with good accuracy.
Comparing Jupiter and Saturn, it is found that the latitudinal shape
of the winds affects considerably the gravity field. The adjoint method
also enables showing which regions of the planet are most impactful
on the gravity field. We find that the gravitational moments are most
sensitive to winds at depths of around $10,000\,\textup{km}$, especially
at the equatorial region, but the signature of deep flows will appear
in the gravity field even if the flows are much shallower. Therefore,
if deep winds exist on these planets, they will likely leave a measurable
signature in the upcoming measurements.

\acknowledgements{Acknowledgments: We thank the Juno science team and Eli Tziperman
for useful discussions. This research has been supported by the Israeli
ministry of Science under grants (3-11481 and 45-851-641), and the
Helen Kimmel Center for Planetary Science at the Weizmann Institute
of Science.}

\appendix

\section{An example on the derivation of the adjoint model\label{Appendix-A}}

In order to give a better understanding of the adjoint method we illustrate
the derivation of the adjoint equations for a simple case. The same
principle could be applied to any set of partial differential equations
(see another example in \citealt{Tziperman1989}). Consider a simple
one-dimensional advection-diffusion equation in steady state for a
tracer $c(x)$
\begin{eqnarray}
uc_{x} & = & kc_{xx},\label{eq:dynamical equation}
\end{eqnarray}
where the parameters we wish to optimize are $u$ and $k$ (the parameters
in this example are equivalent to depth of the wind in our experiments,
and the tracer is equivalent to the density or velocity fields). We
set the two boundary conditions as
\begin{eqnarray}
uc-kc_{x}\vert_{x=0} & = & F_{0},\nonumber \\
uc-kc_{x}\vert_{x=L} & = & F_{L}.
\end{eqnarray}
The cost function is set to be the difference between the calculated
tracer $c$ and the observed tracer $c^{{\rm obs}}$
\begin{eqnarray*}
{\cal J} & = & \frac{1}{2}\int_{0}^{L}\left(c(x)-c^{{\rm obs}}(x)\right)^{2}dx.
\end{eqnarray*}
Next, we define a Langrange function to include the constraints due
to the dynamical equation (\ref{eq:dynamical equation})
\begin{eqnarray}
{\cal L} & = & {\cal J}+\int_{0}^{L}\lambda(uc_{x}-kc_{xx})dx,\label{eq:lagrange equation}
\end{eqnarray}
where $\lambda$ is the Lagrange multiplier that will turn out to
be the adjoint variable for the tracer $c$. As the second term is
identically zero, the values of ${\cal L}$ and ${\cal J}$ are equal.
The minimum of the cost function ${\cal J}$ (model solution) is reached
when the Langrange function ${\cal L}$ has an extremum (zero derivative).
Consider an arbitrary variation in the function $c$,
\begin{eqnarray}
\delta{\cal L=L}(c+\delta c)-{\cal L}(c) & = & \int_{0}^{L}\left(c(x)-c^{obs}(x)\right)\delta cdx+\int_{0}^{L}\lambda(u\delta c_{x}-k\delta c_{xx})dx,
\end{eqnarray}
 where $\delta c$ is an infinitesimally small arbitrary function
in $x$, aside from the boundaries where it must conform to the boundary
conditions (see details below). Applying integration by parts we get
\begin{eqnarray}
\delta{\cal L} & = & \int_{0}^{L}\left[c(x)-c^{{\rm obs}}(x)-(u\lambda_{x}+k\lambda_{xx})\right]\delta cdx\label{eq:lagrange variant}\\
 & + & \lambda\left[u\delta c-k\delta c_{x}\right]_{x=0}^{L}\nonumber \\
 & + & \delta c\left[k\lambda{}_{x}\right]_{x=0}^{L}.\nonumber 
\end{eqnarray}

We need now two conditions under which $\delta{\cal L}$ is zero.
The second line in Eq.~\ref{eq:lagrange variant} has exactly the
formulation of the boundary conditions stated above but for the variation
$\delta c$. Since the boundary conditions should not change with
variations in $c$, this term is zero by definition. Next, we can
demand that the third line vanishes for any $\delta c$, i.e., that
the boundary conditions for the adjoint variable $\lambda$ are $\left[k\lambda{}_{x}\right]_{x=0}^{L}=0$.
We can do so because $\lambda$ is not a physical variable so it's
boundary conditions could be set to fit the requirement on ${\cal L}.$
Finally, we demand that ${\cal \delta L}$ is zero for any function
$\delta c$, therefore the integrand in the first line of Eq.~\ref{eq:lagrange variant}
must vanish, which gives an equation for the adjoint variable $\lambda$
\begin{eqnarray}
u\lambda_{x}+k\lambda_{xx} & = & c(x)-c^{{\rm obs}}(x).\label{eq:adjoint equation}
\end{eqnarray}

Finally, once we have the formulation for the adjoint variable $\lambda$,
we can optimize the cost function with respect to the parameters $u$
and $k$. The derivative of the cost function with respect to these
two parameters could be easily found by differentiation of Eq.~\ref{eq:lagrange equation}
with respect to the two variables, so that
\begin{equation}
\frac{\partial{\cal L}}{\partial u}=\int_{0}^{L}\lambda c_{x}dx,\,\,\,\,\,\,\,\,\,\,\frac{\partial{\cal L}}{\partial k}=-\int_{0}^{L}\lambda c_{xx}dx.\label{eq:parameter-sensitivity}
\end{equation}

Thus, calculating $\lambda$ (the sensitivity of the cost function
with respect to the tracer $c$), and then integrating using Eq.~\ref{eq:parameter-sensitivity},
we can find the gradient of the cost function with respect to the
control variables, i.e., the direction in which those parameters should
be changed in order to reach the minimum of the cost function. Note
that the control variables should be modified in the direction opposite
to the adjoint solution. Given that solving this example, and the
actual model described in this study, needs to be done numerically,
it is important that the adjoint model is actually derived from the
finite difference formulation of the forward model, and not from the
analytical version. The adjoint of the finite difference was shown
to be more accurate than the finite difference of the adjoint \citep{Sirkes1997}.\bibliographystyle{apalike}
\bibliography{bibliography}

\begin{thebibliography}{}

\bibitem[{Atkinson} et~al., 1996]{Atkinson1996}
{Atkinson}, D.~H., {Pollack}, J.~B., and {Seiff}, A. (1996).
\newblock {G}alileo doppler measurements of the deep zonal winds at {J}upiter.
\newblock {\em Science}, 272:842--843.

\bibitem[{Aurnou} et~al., 2008]{Aurnou2008}
{Aurnou}, J., {Heimpel}, M., {Allen}, L., {King}, E., and {Wicht}, J. (2008).
\newblock Convective heat transfer and the pattern of thermal emission on the
  gas giants.
\newblock {\em Geophysical Journal International}, 173:793--801.

\bibitem[{Aurnou} et~al., 2007]{Aurnou2007}
{Aurnou}, J., {Heimpel}, M., and {Wicht}, J. (2007).
\newblock The effects of vigorous mixing in a convective model of zonal flow on
  the ice giants.
\newblock {\em Icarus}, 190:110--126.

\bibitem[Blessing et~al., 2014]{blessing2014testing}
Blessing, S., Kaminski, T., Lunkeit, F., Matei, I., Giering, R., K{\"o}hl, A.,
  Scholze, M., Herrmann, P., Fraedrich, K., and Stammer, D. (2014).
\newblock Testing variational estimation of process parameters and initial
  conditions of an earth system model.
\newblock {\em Tellus A}, 66.

\bibitem[{Bolton}, 2005]{Bolton2005}
{Bolton}, S.~J. (2005).
\newblock {J}uno final concept study report.
\newblock Technical Report AO-03-OSS-03, New Frontiers, NASA.

\bibitem[Busse, 1976]{Busse1976}
Busse, F.~H. (1976).
\newblock A simple model of convection in the {J}ovian atmosphere.
\newblock {\em Icarus}, 29:255--260.

\bibitem[Cho and Polvani, 1996]{Cho1996}
Cho, J. and Polvani, L.~M. (1996).
\newblock The formation of jets and vortices from freely-evolving shallow water
  turbulence on the surface of a sphere.
\newblock {\em Phys. of Fluids.}, 8:1531--1552.

\bibitem[{Del Genio} and {Barbara}, 2012]{DelGenio2012}
{Del Genio}, A. and {Barbara}, J. (2012).
\newblock Constraints on saturn's tropospheric general circulation from cassini
  \{ISS\} images.
\newblock {\em Icarus}, 219(2):689 -- 700.

\bibitem[{Ferreira} et~al., 2005]{Ferreira2005}
{Ferreira}, D., {Marshall}, J., and {Heimbach}, P. (2005).
\newblock Estimating eddy stresses by fitting dynamics to observations using a
  residual-mean ocean circulation model and its adjoint.
\newblock {\em J. Phys. Oceanogr.}, 35:1891.

\bibitem[Finocchiaro, 2013]{Finocchiaro2013}
Finocchiaro, S. (2013).
\newblock Numerical simulations of the {J}uno gravity experiment. {Ph.D.}
  thesis.
\newblock {\em Pubblicazioni Aperte Digitali della Sapienza, code 1889}.

\bibitem[{Finocchiaro} and {Iess}, 2010]{Finocchiaro2010}
{Finocchiaro}, S. and {Iess}, L. (2010).
\newblock Numerical simulations of the gravity science experiment of the {J}uno
  mission to {J}upiter.
\newblock In {\em Spaceflight mechanics}, volume 136, pages 1417--1426. Amer.
  Astro. Soc.

\bibitem[{Galanti} and {Kaspi}, 2015]{Galanti2015DPS}
{Galanti}, E. and {Kaspi}, Y. (2015).
\newblock Deciphering {J}upiter's complex flow dynamics using the upcoming
  {J}uno gravity measurements and an adjoint based dynamical model.
\newblock In {\em AAS/ 47th Division for Planetary Sciences Meeting Abstracts.
  403.08}.

\bibitem[{Galanti} et~al., 2003]{Galanti2003}
{Galanti}, E., {Tziperman}, E., {Harrison}, M., {Rosati}, A., and {Sirkes}, Z.
  (2003).
\newblock A study of {ENSO} prediction using a hybrid coupled model and the
  adjoint method for data assimilation.
\newblock {\em Mon. Weath. Rev.}, 131:2748.

\bibitem[Garcia-Melendo et~al., 2011]{Garcia-Melendo2011}
Garcia-Melendo, E., {Perez-Hoyos}, S., {Sanchez-Lavega}, A., and {Hueso}, R.
  ({2011}).
\newblock {Saturn's zonal wind profile in 2004-2009 from {C}assini {ISS} images
  and its long-term variability}.
\newblock {\em {Icarus}}, {215}({1}):{62--74}.

\bibitem[{Gastine} et~al., 2013]{Gastine2013}
{Gastine}, T., {Wicht}, J., and {Aurnou}, J.~M. (2013).
\newblock Zonal flow regimes in rotating anelastic spherical shells: An
  application to giant planets.
\newblock {\em Icarus}, 225:156--172.

\bibitem[{Heimpel} and {Aurnou}, 2007]{Heimpel2007}
{Heimpel}, M. and {Aurnou}, J. (2007).
\newblock Turbulent convection in rapidly rotating spherical shells: A model
  for equatorial and high latitude jets on {J}upiter and {S}aturn.
\newblock {\em Icarus}, 187:540--557.

\bibitem[Heimpel et~al., 2015]{Heimpel2015}
Heimpel, M., Gastine, T., and Wicht, J. (2015).
\newblock Simulation of deep-seated zonal jets and shallow vortices in gas
  giant atmospheres.
\newblock {\em Nature Geosci}, advance online publication.

\bibitem[Hestenes, 1980]{Herstenes1980}
Hestenes, M. (1980).
\newblock {\em Conjugate direction methods in optimization}.

\bibitem[Hubbard, 1982]{Hubbard1982}
Hubbard, W. ({1982}).
\newblock {Eeffects Of Differential Rotations On The Gravitational Figures Of
  Jupiter And Saturn}.
\newblock {\em {Icarus}}, {52}({3}):{509--515}.

\bibitem[{Hubbard}, 1984]{Hubbard1984}
{Hubbard}, W.~B. (1984).
\newblock {\em Planetary interiors}.
\newblock New York, Van Nostrand Reinhold Co., 1984, 343 p.

\bibitem[{Hubbard}, 1999]{Hubbard1999}
{Hubbard}, W.~B. (1999).
\newblock Note: Gravitational signature of {J}upiter's deep zonal flows.
\newblock {\em Icarus}, 137:357--359.

\bibitem[{Hubbard}, 2012]{Hubbard2012}
{Hubbard}, W.~B. (2012).
\newblock High-precision {M}aclaurin-based models of rotating liquid planets.
\newblock {\em Astrophys. J. Let.}, 756:L15.

\bibitem[Hubbard, 2013]{Hubbard2013}
Hubbard, W.~B. (2013).
\newblock Conventric maclaurian spheroid models of rotating liquid planets.
\newblock {\em Astrophys. J.}, 768(1).

\bibitem[{Hubbard} et~al., 2014]{Hubbard2014}
{Hubbard}, W.~B., {Schubert}, G., {Kong}, D., and {Zhang}, K. (2014).
\newblock On the convergence of the theory of figures.
\newblock {\em Icarus}, 242:138--141.

\bibitem[Kalmikov and Heimbach, 2014]{Kalmikov2014}
Kalmikov, A. and Heimbach, P. (2014).
\newblock A hessian-based method for uncertainty quantification in global ocean
  state estimation.
\newblock {\em SIAM Journal on Scientific Computing}, 36:S267--95.

\bibitem[{Kaspi}, 2013]{Kaspi2013a}
{Kaspi}, Y. (2013).
\newblock Inferring the depth of the zonal jets on {J}upiter and {S}aturn from
  odd gravity harmonics.
\newblock {\em Geophys. Res. Lett.}, 40:676--680.

\bibitem[{Kaspi} et~al., 2013a]{Kaspi2013d}
{Kaspi}, Y., {Davighi}, J., {Galanti}, E., and {Hubbard}, W. (2013a).
\newblock Estimating the depth of the zonal jet streams on {J}upiter and
  {S}aturn through inversion of gravity measurements by {J}uno and {C}assin.
\newblock {\em AGU Fall Meeting Abstracts}.

\bibitem[{Kaspi} et~al., 2016]{Kaspi2016}
{Kaspi}, Y., {Davighi}, J.~E., {Galanti}, E., and {Hubbard}, W.~B. (2016).
\newblock The gravity signature of atmospheric dynamics on giant planets:
  comparison between the potential theory and thermal wind approaches.
\newblock {\em Icarus}.
\newblock under revision.

\bibitem[Kaspi and Flierl, 2007]{Kaspi2007}
Kaspi, Y. and Flierl, G.~R. (2007).
\newblock Formation of jets by baroclinic instability on gas planet
  atmospheres.
\newblock {\em J. Atmos. Sci.}, 64:3177--3194.

\bibitem[{Kaspi} et~al., 2009]{Kaspi2009}
{Kaspi}, Y., {Flierl}, G.~R., and {Showman}, A.~P. (2009).
\newblock The deep wind structure of the giant planets: Results from an
  anelastic general circulation model.
\newblock {\em Icarus}, 202:525--542.

\bibitem[{Kaspi} et~al., 2010]{Kaspi2010a}
{Kaspi}, Y., {Hubbard}, W.~B., {Showman}, A.~P., and {Flierl}, G.~R. (2010).
\newblock Gravitational signature of {J}upiter's internal dynamics.
\newblock {\em Geophys. Res. Lett.}, 37:L01204.

\bibitem[{Kaspi} et~al., 2013b]{Kaspi2013c}
{Kaspi}, Y., {Showman}, A.~P., {Hubbard}, W.~B., {Aharonson}, O., and {Helled},
  R. (2013b).
\newblock Atmospheric confinement of jet streams on {U}ranus and {N}eptune.
\newblock {\em Nature}, 497:344--347.

\bibitem[{Kong} et~al., 2012]{Kong2012}
{Kong}, D., {Zhang}, K., and {Schubert}, G. (2012).
\newblock On the variation of zonal gravity coefficients of a giant planet
  caused by its deep zonal flows.
\newblock 748.

\bibitem[{Li} et~al., 2006]{Li2006}
{Li}, L., {Ingersoll}, A., {Vasavada}, A., {Simon-Miller}, A., {Del Genio}, A.,
  {Ewald}, S., {Porco}, C., and {West}, R. (2006).
\newblock Vertical wind shear on jupiter from cassini images.
\newblock {\em Journal of Geophysical Research: Planets}, 111(E4):n/a--n/a.

\bibitem[{Lian} and {Showman}, 2010]{Lian2010}
{Lian}, Y. and {Showman}, A.~P. (2010).
\newblock Generation of equatorial jets by large-scale latent heating on the
  giant planets.
\newblock {\em Icarus}, 207:373--393.

\bibitem[{Liu} et~al., 2008]{Liu2008}
{Liu}, J., {Goldreich}, P.~M., and {Stevenson}, D.~J. (2008).
\newblock Constraints on deep-seated zonal winds inside {J}upiter and {S}aturn.
\newblock {\em Icarus}, 196:653--664.

\bibitem[{Liu} and {Schneider}, 2010]{Liu2010}
{Liu}, J. and {Schneider}, T. (2010).
\newblock Mechanisms of jet formation on the giant planets.
\newblock {\em J. Atmos. Sci.}, 67:3652--3672.

\bibitem[Liu et~al., 2014]{Liu2014}
Liu, J., Schneider, T., and Fletcher, L.~N. (2014).
\newblock Constraining the depth of saturn's zonal winds by measuring thermal
  and gravitational signals.
\newblock {\em Icarus}, 239:260--272.

\bibitem[{Liu} et~al., 2013]{Liu2013}
{Liu}, J., {Schneider}, T., and {Kaspi}, Y. (2013).
\newblock Predictions of thermal and gravitational signals of {J}upiter's deep
  zonal winds.
\newblock {\em Icarus}, 224:114--125.
\newblock under revision.

\bibitem[Marotzke et~al., 1999]{Marotzke1999}
Marotzke, J., {Giering}, R., {Zhang}, K., {Stammer}, D., {Hill}, C., and {Lee},
  T. ({1999}).
\newblock {Construction of the adjoint {MIT} ocean general circulation model
  and application to {A}tlantic heat transport sensitivity}.
\newblock {\em {J. Geophys. Res. - Oceans}}, {104}({C12}):{29529--29547}.

\bibitem[{Mazloff} et~al., 2010]{Mazloff2010}
{Mazloff}, M.~R., {Heimbach}, P., and {Wunsch}, C. (2010).
\newblock An eddy-permitting southern ocean state estimate.
\newblock {\em J. Phys. Oceanogr.}, 40:880--899.

\bibitem[{Moore} et~al., 2011]{Moore2011}
{Moore}, A.~M., {Arango}, H.~G., {Broquet}, G., {Powell}, B.~S., {Weaver},
  A.~T., and {Zavala-Garay}, J. (2011).
\newblock The regional ocean modeling system ({ROMS}) 4-dimensional variational
  data assimilation systems . part {I} - system overview and formulation.
\newblock {\em Prog. Oceanogr.}, 91:34--49.

\bibitem[{Parisi} et~al., 2016]{Parisi2016}
{Parisi}, M., {Galanti}, E., {Finocchiaro}, S., {Iess}, L., and {Kaspi}, Y.
  (2016).
\newblock Probing the atmospheric dynamics of {J}upiter's {G}reat {R}ed {S}pot
  with the {J}uno gravity experiment.
\newblock {\em Icarus, in press}.

\bibitem[Pedlosky, 1987]{Pedlosky1987}
Pedlosky, J. (1987).
\newblock {\em Geophysical Fluid Dynamics}.
\newblock Spinger.

\bibitem[{Porco} et~al., 2003]{Porco2003}
{Porco}, C.~C., {West}, R.~A., {McEwen}, A., {Del Genio}, A.~D., {Ingersoll},
  A.~P., {Thomas}, P., {Squyres}, S., {Dones}, L., {Murray}, C.~D., {Johnson},
  T.~V., {Burns}, J.~A., {Brahic}, A., {Neukum}, G., {Veverka}, J., {Barbara},
  J.~M., {Denk}, T., {Evans}, M., {Ferrier}, J.~J., {Geissler}, P.,
  {Helfenstein}, P., {Roatsch}, T., {Throop}, H., {Tiscareno}, M., and
  {Vasavada}, A.~R. (2003).
\newblock {C}assini imaging of {J}upiter's atmosphere, satellites and rings.
\newblock {\em Science}, 299:1541--1547.

\bibitem[{Sanchez-Lavega} et~al., 2007]{SanchezLavega2007}
{Sanchez-Lavega}, A., {Hueso}, R., and {Perez-Hoyos}, S. (2007).
\newblock The three-dimensional structure of saturn's equatorial jet at cloud
  level.
\newblock {\em Icarus}, 187(2):510 -- 519.

\bibitem[{Sanchez-Lavega} et~al., 2000]{Sanchez-Lavega2000}
{Sanchez-Lavega}, A., {Rojas}, J.~F., and {Sada}, P.~V. (2000).
\newblock Saturn's zonal winds at cloud level.
\newblock {\em Icarus}, 147:405--420.

\bibitem[{Schneider} and {Liu}, 2009]{Schneider2009}
{Schneider}, T. and {Liu}, J. (2009).
\newblock Formation of jets and equatorial superrotation on {J}upiter.
\newblock {\em J. Atmos. Sci.}, 66:579--601.

\bibitem[{Scott} and {Polvani}, 2007]{Scott2007}
{Scott}, R.~K. and {Polvani}, L.~M. (2007).
\newblock Forced-dissipative shallow-water turbulence on the sphere and the
  atmospheric circulation of the giant planets.
\newblock {\em J. Atmos. Sci.}, 64:3158--3176.

\bibitem[{Showman} et~al., 2006]{Showman2006}
{Showman}, A.~P., {Gierasch}, P.~J., and {Lian}, Y. (2006).
\newblock Deep zonal winds can result from shallow driving in a giant-planet
  atmosphere.
\newblock {\em Icarus}, 182:513--526.

\bibitem[Sirkes and Tziperman, 1997]{Sirkes1997}
Sirkes, Z. and Tziperman, E. ({1997}).
\newblock {Finite difference of adjoint or adjoint of finite difference?}
\newblock {\em {Monthly Weather Review}}, {125}({12}):{3373--3378}.

\bibitem[Thacker and {Long}, 1988]{Tacker1988}
Thacker, W. and {Long}, R. ({1988}).
\newblock {Fitting Dynamics To Ddata}.
\newblock {\em {J. Geophys. Res. - Oceans}}, {93}({C2}):{1227--1240}.

\bibitem[{Tziperman}, 1992]{Tziperman1992}
{Tziperman}, E. (1992).
\newblock Computing the steady oceanic circulation using an optimization
  approach.
\newblock {\em Dyn. Atmos. Oceans}, 16:379--403.

\bibitem[{Tziperman} and {Thacker}, 1989]{Tziperman1989}
{Tziperman}, E. and {Thacker}, W.~C. (1989).
\newblock An optimal-control/adjoint-equations approach to studying the oceanic
  general circulation.
\newblock {\em J. Phys. Oceanogr.}, 19:1471--1485.

\bibitem[{Vasavada} and {Showman}, 2005]{Vasavada2005}
{Vasavada}, A.~R. and {Showman}, A.~P. (2005).
\newblock {J}ovian atmospheric dynamics: {A}n update after {G}alileo and
  {C}assini.
\newblock {\em Reports of Progress in Physics}, 68:1935--1996.

\bibitem[Williams, 1978]{Williams1978}
Williams, G.~P. (1978).
\newblock Planetary circulations: 1. barotropic representation of the {J}ovian
  and terrestrial turbulence.
\newblock {\em J. Atmos. Sci.}, 35:1399--1426.

\bibitem[Wisdom and Hubbard, 2016]{Wisdom2016}
Wisdom, J. and Hubbard, W. ({2016}).
\newblock {Differential Rotation in {J}upiter: a comparison of Methods}.
\newblock {\em {Icarus, submitted}}.

\bibitem[{Wunsch} and {Heimbach}, 2007]{Wunsch2007}
{Wunsch}, C. and {Heimbach}, P. (2007).
\newblock Practical global oceanic state estimation.
\newblock {\em Physica D Nonlinear Phenomena}, 230:197--208.

\bibitem[{Zhang} et~al., 2015]{Zhang2015}
{Zhang}, K., {Kong}, D., and {Schubert}, G. ({2015}).
\newblock {Thermal-gravitational wind equation for the wind-induced
  gravitational signature of giant gaseous planets: mathematical derivation,
  numerical method, and illustrative solutions}.
\newblock {\em Astrophys. J.}, {806}({2}).

\bibitem[{Zharkov} and {Trubitsyn}, 1978]{Zharkov1978}
{Zharkov}, V.~N. and {Trubitsyn}, V.~P. (1978).
\newblock {\em Physics of planetary interiors}.
\newblock pp.~388.~Pachart Publishing House.

\end{thebibliography}

\end{document}